\documentclass[aps,prd,preprintnumbers,groupedaddress,nofootinbib,amssymb,eqsecnum,longbibliography,10pt]{revtex4-2}

\usepackage{graphicx}
\usepackage{color} 
\usepackage{dcolumn}
\usepackage{amsmath,amsthm,amssymb}
\usepackage{amsfonts}
\usepackage{bm}
\usepackage{mathrsfs}
\usepackage{revsymb}
\usepackage{comment}
\usepackage{braket}
\usepackage{tikz}
\usetikzlibrary{decorations.pathmorphing,arrows.meta,calc}
\usetikzlibrary{
  positioning,
  decorations.pathmorphing,
  arrows.meta
}

\usepackage[bookmarks=true,bookmarksnumbered=true,colorlinks=false]{hyperref}
\newif\ifoneauthor
\oneauthortrue

\begin{document}

\title{Gravitationally Induced Quantum Decoherence of Macroscopic Objects}

\author{Hiroki Takeda}
\email[]{takeda@tap.scphys.kyoto-u.ac.jp}
\affiliation{The Hakubi Center for Advanced Research, Kyoto University, Kyoto 606-8501, Japan}
\affiliation{Department of Physics, Kyoto University, Kyoto 606-8502, Japan}

\author{Shogo Tomizuka}
\affiliation{Department of Physics, Kyoto University, Kyoto 606-8502, Japan}

\author{Takahiro Tanaka}
\affiliation{Department of Physics, Kyoto University, Kyoto 606-8502, Japan}
\affiliation{Center for Gravitational Physics and Quantum Information, Yukawa Institute for Theoretical Physics, Kyoto University, Kyoto 606-8502, Japan}

\date{\today}

\begin{abstract}
We formulate the gravitationally induced quantum decoherence of a massive object prepared in a spatial superposition. 
Starting from linearized gravity coupled to a massive system particle and an environmental scalar field, we derive a closed-time-path influence functional governing the reduced system dynamics. 
In the nonrelativistic and quasi-static regime, the decoherence exponent can be written as a bilinear functional of the difference of the system stress–energy tensors and an effective noise kernel obtained by dressing the environmental stress–energy tensor correlator with graviton propagators. 
We then apply this framework to the Newtonian long-range gravitational interaction and evaluate the resulting decoherence function for a dilute nonrelativistic gas modeled by finite wave packets and coarse-grained in time and space. 
By performing controlled approximations, we obtain analytic expressions for the cumulative decoherence function and show that the dominant contribution is accumulated logarithmically over a broad range of distances, while remaining subdominant to conventional collisional decoherence under realistic conditions.
\end{abstract}

\maketitle


\section{Introduction}
Macroscopic quantum superpositions of massive objects play a central role in several frontiers of quantum science.
Matter-wave interferometry, levitated particles, and optomechanical systems are extending quantum control toward larger masses and longer length scales~\cite{Cronin:2009zz, Aspelmeyer:2013lha, Arndt:2014blv, Millen:2019bcw}.
These platforms are also being explored for quantum-enhanced sensing~\cite{Degen:2016pxo}, dark matter searches~\cite{Carney:2020xol}, and tests of the quantum nature of gravity~\cite{Bose:2017nin, Marletto:2017kzi}.
In all of these settings, a central question is how rapidly coherence is lost once a macroscopic object is coupled to its environment.

Quantum decoherence is the loss of phase coherence between different branches of a quantum superposition induced by entanglement with environmental degrees of freedom~\cite{Zeh:1970zz, Zurek:1981xq, Zurek:1982ii, Giulini:1996nw, Zurek:2003zz, Schlosshauer:2003zy, Schlosshauer2007, Schlosshauer:2019ewh}.
For ordinary environmental channels, decoherence is usually dominated by electromagnetic noise, thermal emission, or collisions with residual gas particles.
In particular, residual-gas decoherence has been studied extensively within collisional models formulated in terms of scattering amplitudes and cross sections~\cite{Joos:1984uk, Gallis:1990bcf, Hornberger:2003umw}. 

Gravity, however, raises a conceptually distinct issue. 
Although gravitational interactions are extremely weak at the microscopic level, they are long ranged, universal, and essentially impossible to screen.
A key question is whether a dilute gas can induce an additional decoherence channel through the gravitational interaction.
The standard collisional framework is not directly applicable in this context, because the Newtonian interaction is long ranged and leads to infrared-enhanced small-angle scattering that cannot be naturally reduced to a description in terms of localized collision events and finite scattering cross sections.

In this paper, we formulate the gravitationally induced decoherence of a macroscopic object from first principles using the closed-time-path framework of open quantum systems~\cite{Feynman:1963fq,Schwinger:1960qe, Keldysh:1964ud, Calzetta:2008iqa}. 
Starting from linearized gravity coupled to a massive system particle and a scalar environment, we integrate out the environmental degrees of freedom and derive a decoherence functional in terms of the difference of the system stress--energy tensors and an effective gravitationally dressed noise kernel. 
The resulting formalism provides a unified description of gravitationally induced decoherence arising from environmental matter fluctuations as well as intrinsic graviton fluctuations~\cite{Anastopoulos:2013zya, Blencowe:2012mp}.

We then apply this framework to the channel most relevant for present laboratory systems, namely the Newtonian gravitational interaction between the superposed object and a dilute nonrelativistic gas. 
Modeling the gas by finite wave packets and coarse graining in time and space, we introduce a cumulative spatial-window decoherence function whose diagonal contribution reduces to a one-dimensional momentum integral.
This formulation makes it possible to track the radial buildup of decoherence, isolate the logarithmically enhanced interval, and evaluate the saturated contribution for experimentally motivated parameters.
We show that the logarithmic enhancement is accumulated over a broad range of distances rather than a localized region.
Under realistic gas conditions, however, the gravitational contribution remains subdominant to standard collisional decoherence. 
These results provide a quantitative benchmark for identifying parameter regimes in which observable gravitational decoherence could arise.

Throughout this paper, we adopt natural units where $c=\hbar=1$ unless explicitly restored.
We define the gravitational coupling $\kappa$ and the reduced Planck mass $M_{\rm pl}$ by $\kappa^2 = 32\pi G$ and $M_{\rm pl}^2 = {1}/(8\pi G)$, respectively.
We use the metric signature $(-,+,+,+)$.
We write $x := (x^0, {\bm x})$ for spacetime coordinates, and $k := (k^0, {\bm k})$ for the corresponding four-momentum.
For any tensor component $T_{\#}(x)$, we define the Fourier transform by $T_{\#}(x)={1}/{(2\pi)^4}\int d^4k\; \widetilde{T}_{\#}(k)\, e^{ik_\mu x^\mu}$.
For a two-point kernel $T_{\#}(x,y)$, we similarly use the double Fourier transform $T_{\#}(x,y) = 1/(2\pi)^8 \int d^4k\,d^4k'\; e^{ik_\mu x^{\mu}}\widetilde T_{\#}(k,k')  e^{-ik'_\mu y^{\mu}}$.

\section{Closed-time-path formulation}
\label{sec:CTP}
In this section we formulate the reduced dynamics of a massive system particle interacting with an environmental scalar field through linearized gravity using the closed time path formalism. 
Our aim is to derive an influence functional governing the system worldline after the metric perturbation and the scalar environment have been traced out. 
To this end we first specify the microscopic action of the coupled system, then expand the metric around Minkowski spacetime and fix the gauge, and finally rewrite the reduced density matrix in terms of an influence functional describing the reduced system dynamics.

\subsection{Microscopic model}
\label{subsec:CTP_model}

We consider a microscopic total action of the form
\begin{align}
S_{\rm tot}[g_{\mu\nu},X^{\mu},\Phi]
=
\int d^4x\,\mathcal L_{\rm tot}(g_{\mu\nu},X^{\mu},\Phi),
\end{align}
where the total Lagrangian density is decomposed into gravitational and matter parts,
\begin{align}
\mathcal L_{\rm tot}(g_{\mu\nu},X^{\mu},\Phi)
=
\mathcal L_{\rm EH}(g_{\mu\nu})
+
\mathcal L_m(g_{\mu\nu},X^{\mu},\Phi).
\end{align}
The gravitational sector is described by the Einstein--Hilbert action
\begin{align}
S_{\rm EH}[g_{\mu\nu}]
=
\int d^4x\,\mathcal L_{\rm EH}(g_{\mu\nu})
=
\int d^4x\,\sqrt{-g}\,\frac{1}{16\pi G}\,R,
\label{eq:EH_action}
\end{align}
where $g=\det(g_{\mu\nu})$ and $R$ is the Ricci scalar.

The matter sector is divided into the system degree of freedom and the environment,
\begin{align}
S_m[g_{\mu\nu},X^{\mu},\Phi]
=
\int d^4x\,\mathcal L_m(g_{\mu\nu},X^{\mu},\Phi)
=
S_{\rm sys}[g_{\mu\nu},X^{\mu}]
+
S_{\rm env}[g_{\mu\nu},\Phi].
\end{align}
The system variable $X^\mu$ denotes the spacetime trajectory of a point particle of mass $M$.
$X^\mu(\lambda)$ is a map from a one-dimensional parameter space into spacetime, where $\lambda$ is an arbitrary worldline parameter and $\tau$ is the proper time along the trajectory.
The system action is therefore
\begin{align}
S_{\rm sys}[g_{\mu\nu},X^{\mu}]
=
- M\int d\tau
=
- M\int d\lambda\,
\sqrt{
- g_{\mu\nu}\bigl(X(\lambda)\bigr)
\frac{dX^\mu}{d\lambda}
\frac{dX^\nu}{d\lambda}
}.
\label{eq:Ssys}
\end{align}

The environment is taken to be a complex Klein--Gordon scalar field $\Phi$ of mass $m$.
Its action is
\begin{align}
S_{\rm env}[g_{\mu\nu},\Phi]
=
\int d^4x\,\sqrt{-g}
\Big(
- g^{\mu\nu}\partial_\mu\Phi^*\partial_\nu\Phi
- m^2\Phi^*\Phi
\Big),
\label{eq:Senv}
\end{align}
where $\partial_\mu=\partial/\partial x^\mu$.

\subsection{Metric perturbation and gauge fixing}
\label{subsec:CTP_gauge}

We expand the metric around Minkowski spacetime as
\begin{align}
g_{\mu\nu}
=
\eta_{\mu\nu}
+
\kappa h_{\mu\nu}.
\label{eq:metric_perturb}
\end{align}
Substituting this into the total action and keeping terms up to quadratic order in $h_{\mu\nu}$, we obtain
\begin{align}
S_{\rm tot}[h_{\mu\nu},X^\mu,\Phi]
=
S_{\rm EH}^{(2)}[h_{\mu\nu}]
+
S_{\rm sys}^{(0)}[X^\mu]
+
S_{\rm env}^{(0)}[\Phi]
+
S_{\rm int}[h_{\mu\nu},X^\mu,\Phi].
\label{eq:Stot_quad}
\end{align}
The quadratic gravitational action is
\begin{align}
S_{\rm EH}^{(2)}[h_{\mu\nu}]
=
\frac12\int d^4x\;
h_{\mu\nu}\mathcal E^{\mu\nu\alpha\beta}h_{\alpha\beta},
\label{eq:SEH2}
\end{align}
where $\mathcal E^{\mu\nu\alpha\beta}$ denotes the linearized Einstein--Hilbert operator.

The flat-background system and environment actions are
\begin{align}
S_{\rm sys}^{(0)}[X^\mu]
&=
- M\int d\lambda\,
\sqrt{-\eta_{\mu\nu} \frac{d X^\mu}{d\lambda}\frac{d X^\nu}{d\lambda}},
\label{eq:Ssys0}
\\
S_{\rm env}^{(0)}[\Phi]
&=
\int d^4x\,
\Big(
-\eta_{\mu\nu}\partial^\mu\Phi^*\partial^\nu\Phi
-m^2\Phi^*\Phi
\Big).
\label{eq:Senv0}
\end{align}
The interaction with the metric perturbation is linear in $h_{\mu\nu}$ and is given by a coupling to the total stress–energy tensor,
\begin{align}
S_{\rm int}[h_{\mu\nu},X^\mu,\Phi]
&=
S_{\rm int}^{\rm sys}[h_{\mu\nu},X^\mu]
+
S_{\rm int}^{\rm env}[h_{\mu\nu},\Phi]
\nonumber\\
&=
\frac{\kappa}{2}
\int d^4x\;
h_{\mu\nu}
\Big(
T_{\rm sys}^{\mu\nu}+T_{\rm env}^{\mu\nu}
\Big).
\label{eq:Sint_linear}
\end{align}
The stress--energy tensor of the point particle is
\begin{align}
T_{\rm sys}^{\mu\nu}(x)
=
M\int d\lambda\;
\delta^{(4)}\!\bigl(x-X(\lambda)\bigr)\,
\frac{\dfrac{dX^\mu}{d\lambda}\dfrac{dX^\nu}{d\lambda}}
{\sqrt{-\eta_{\rho\sigma}\dfrac{dX^\rho}{d\lambda}\dfrac{dX^\sigma}{d\lambda}}},
\label{eq:Tsys}
\end{align}
while that of the scalar field is
\begin{align}
T_{\rm env}^{\mu\nu}(x)
=
\partial^\mu\Phi^*\partial^\nu\Phi
+
\partial^\nu\Phi^*\partial^\mu\Phi
-
\eta^{\mu\nu}
\Big(
\partial_\alpha\Phi^*\partial^\alpha\Phi
+
m^2\Phi^*\Phi
\Big).
\label{eq:Tenv}
\end{align}

To quantize $h_{\mu\nu}$, we must remove the residual linearized gauge redundancy.
The action $S_{\rm EH}^{(2)}[h_{\mu\nu}]$ is invariant under
\begin{align}
h_{\mu\nu}\to h_{\mu\nu}-\partial_{(\mu}\xi_{\nu)},
\end{align}
and the corresponding kinetic operator is therefore not invertible.
We impose the de~Donder gauge condition
\begin{align}
\partial^\mu\bar h_{\mu\nu}=0,
\qquad
\bar h_{\mu\nu}=h_{\mu\nu}-\tfrac12\eta_{\mu\nu}h,
\qquad
h=\eta^{\mu\nu}h_{\mu\nu}.
\label{eq:deDonder}
\end{align}
The associated gauge-fixing and Faddeev--Popov ghost actions are
\begin{align}
S_{\rm gf}[h]
&=
-\frac12\int d^4x\;
(\partial^\mu\bar h_{\mu\nu})(\partial_\rho\bar h^{\rho\nu}),
\label{eq:S_gf}
\\
S_{\rm gh}[c,\bar c]
&=
\int d^4x\;
\bar c_\nu(-\Box)c^\nu.
\label{eq:S_gh}
\end{align}
At quadratic order the ghost fields are free and do not couple to the matter sector.
Their path integrals therefore contribute only an overall normalization, and will be omitted hereafter.

We combine the quadratic Einstein--Hilbert action and the gauge-fixing term into
\begin{align}
S_{\rm grav}^{(0)}[h]
:=
S_{\rm EH}^{(2)}[h]
+
S_{\rm gf}[h].
\label{eq:Sgrav0_def}
\end{align}
The gauge-fixed kinetic operator is now invertible, and the corresponding graviton Feynman propagator is
\begin{align}
G_{\mu\nu\alpha\beta}^{F}(k)
=
\frac{i\hbar\,P_{\mu\nu\alpha\beta}}{k^2+i\epsilon},
\label{eq:graviton_Feynman}
\end{align}
where $P_{\mu\nu\alpha\beta}
:=
(1/2)
(
\eta_{\mu\alpha}\eta_{\nu\beta}
+
\eta_{\mu\beta}\eta_{\nu\alpha}
-
\eta_{\mu\nu}\eta_{\alpha\beta}
)$.

\subsection{Influence functional}
\label{subsec:CTP_IF}
Let $t_i$ and $t_f$ ($t_i<t_f$) be fixed parameters specifying the initial and final times of the field variables, respectively.
We introduce the closed time contour
$\mathcal C: t_i \rightarrow t_f \rightarrow t_i$.
For each dynamical variable $Q\in\{h_{\mu\nu},X^\mu,\Phi\}$ we introduce two copies
$Q_+$ and $Q_-$ defined on the forward and backward branches of $\mathcal C$.
The total density operator evolves unitarily according to
\begin{align}
\rho_{\mathrm{tot}}(t_f)
=
U(t_f,t_i)\,\rho_i^{\mathrm{tot}}\,U^\dagger(t_f,t_i),
\qquad
U(t_f,t_i)
=
\mathrm T\exp\!\left[-\frac{i}{\hbar}\int_{t_i}^{t_f}dt\,H(t)\right],
\label{eq:rho_tot_CTP}
\end{align}
where $\rho_i^{\mathrm{tot}}$ denotes the initial density operator of the full system on the Hilbert space of $\{h_{\mu\nu},X^\mu,\Phi\}$.
The reduced density operator of the system is then defined by tracing out the gravitational and scalar environmental degrees of freedom,
\begin{align}
\rho_r(t_f):=\mathrm{Tr}_{h,\Phi}\,\rho_{\mathrm{tot}}(t_f).
\label{eq:rho_r_def}
\end{align}

We assume that the initial state factorizes as
\begin{align}
\rho_i^{\mathrm{tot}}
=
\rho_i^{(h)}
\otimes
\rho_i^{(X)}
\otimes
\rho_i^{(\Phi)},
\label{eq:rho_factorized_CTP}
\end{align}
where $\rho_i^{(h)}$, $\rho_i^{(X)}$, and $\rho_i^{(\Phi)}$ denote the initial density operators
of the gravitational field $h_{\mu\nu}$, the system worldline degree of freedom $X^\mu$, and the scalar field $\Phi$, respectively.

This assumption allows all dependence on the environmental variables to be collected into a single influence functional.
In the $X^\mu$ representation, the reduced density matrix can then be written as
\begin{align}
\rho_r[X_{f,+}^{\mu},X_{f,-}^{\mu};t_f]
&=
\int dX_{i,+}^{\mu}\,dX_{i,-}^{\mu}\,
\rho_i^{(X)}[X_{i,+}^{\mu},X_{i,-}^{\mu}]
\int_{X_+(t_i)=X_{i,+}}^{X_+(t_f)=X_{f,+}}
\mathcal D X_+(t)
\int_{X_-(t_i)=X_{i,-}}^{X_-(t_f)=X_{f,-}}
\mathcal D X_-(t)
\nonumber\\
&\quad\times
\exp\!\Bigg[
\frac{i}{\hbar}
\Big(
S_{\rm sys}^{(0)}[X_+]-S_{\rm sys}^{(0)}[X_-]
\Big)
\Bigg]
\mathcal F[X_\pm],
\label{eq:rho_r_with_IF}
\end{align}
where the influence functional is defined by
\begin{align}
\mathcal F[X_\pm]
:=
\exp\!\left(
\frac{i}{\hbar}S_{\rm IF}[X_\pm]
\right).
\label{eq:IF_def}
\end{align}
Here, $\mathcal F[X_\pm]$ is obtained by integrating the doubled fields $h_\pm$ and $\Phi_\pm$ over the closed time contour,
\begin{align}
\mathcal F[X_\pm]
&=
\int Dh_i^{+}({\bm x})\,Dh_i^{-}({\bm x})\,
D\Phi_i^{+}({\bm x})\,D\Phi_i^{-}({\bm x})\,
\rho_i^{(h)}[h_i^{+},h_i^{-}]\,
\rho_i^{(\Phi)}[\Phi_i^{+},\Phi_i^{-}]
\nonumber\\
&\quad\times
\int_{\,h_{+}(t_i,{\bm x})=h_i^{+}({\bm x})}^{\,h_{+}(t_f,{\bm x})}
\mathcal D h_{+\,\mu\nu}(t,{\bm x})
\int_{\,h_{-}(t_i,{\bm x})=h_i^{-}({\bm x})}^{\,h_{-}(t_f,{\bm x})}
\mathcal D h_{-\,\mu\nu}(t,{\bm x})
\nonumber\\
&\quad\times
\int_{\,\Phi_{+}(t_i,{\bm x})=\Phi_i^{+}({\bm x})}^{\,\Phi_{+}(t_f,{\bm x})}
\mathcal D \Phi_{+}(t,{\bm x})
\int_{\,\Phi_{-}(t_i,{\bm x})=\Phi_i^{-}({\bm x})}^{\,\Phi_{-}(t_f,{\bm x})}
\mathcal D \Phi_{-}(t,{\bm x})
\nonumber\\
&\quad\times
\delta[h_+(t_f,{\bm x})-h_-(t_f,{\bm x})]\,
\delta[\Phi_+(t_f,{\bm x})-\Phi_-(t_f,{\bm x})]
\nonumber\\
&\quad\times
\exp\!\Bigg\{
\frac{i}{\hbar}
\Big(
S_{\rm grav}^{(0)}[h_+]-S_{\rm grav}^{(0)}[h_-]
+
S_{\rm env}^{(0)}[\Phi_+]-S_{\rm env}^{(0)}[\Phi_-]
+
S_{\rm int}[h_+,X_+,\Phi_+]
-
S_{\rm int}[h_-,X_-,\Phi_-]
\Big)
\Bigg\}.
\label{eq:IF_path_def}
\end{align}
This makes explicit that all environmental effects on the system worldline are encoded in the influence functional.

It is useful to introduce the Keldysh combinations of the system stress--energy tensor,
\begin{align}
T^{\rm sys}_{c\,\mu\nu}(x)
:=
\frac12
\Big(
T^{\rm sys}_{+\,\mu\nu}(x)+T^{\rm sys}_{-\,\mu\nu}(x)
\Big),
\qquad
T^{\rm sys}_{\Delta\,\mu\nu}(x)
:=
T^{\rm sys}_{+\,\mu\nu}(x)-T^{\rm sys}_{-\,\mu\nu}(x).
\label{eq:Keldysh_Tsys}
\end{align}
The combination $T_c^{\rm sys}$ describes the average history, while $T_\Delta^{\rm sys}$ measures the difference between the two branches and therefore controls the suppression of off-diagonal elements of the reduced density matrix.

In Eq.~\eqref{eq:IF_path_def}, the environmental degrees of freedom consist of the metric perturbation $h_{\mu\nu}$ and the scalar field $\Phi$.
Since the metric perturbation appears quadratically in $S_{\rm grav}^{(0)}$ and linearly in the interaction, it can be integrated out exactly, yielding a Gaussian influence action for the total stress--energy tensor.
The remaining scalar field $\Phi$ is then traced out assuming a Gaussian initial state.
Using the cumulant expansion for the environmental stress--energy tensor and truncating the connected correlators at second order, we obtain the influence action
\begin{align}
\frac{i}{\hbar}S_{\rm IF}[X_\pm]
&=
-\frac{1}{2\hbar^2}\frac{\kappa^2}{4}
\int d^4x\,d^4y\,
\Big[
-2i\,
T_{\Delta}^{\rm sys\,\mu\nu}(x)
\mathcal D^{(h)}_{\mu\nu\alpha\beta}(x,y)
T_c^{\rm sys\,\alpha\beta}(y)
\nonumber\\
&\qquad
+\,
T_{\Delta}^{\rm sys\,\mu\nu}(x)
\mathcal N^{(h)}_{\mu\nu\alpha\beta}(x,y)
T_{\Delta}^{\rm sys\,\alpha\beta}(y)
-2i\,
T_{\Delta}^{\rm sys\,\mu\nu}(x)
\mathcal D^{(h)}_{\mu\nu\alpha\beta}(x,y)
\big\langle T_c^{{\rm env}\,\alpha\beta}(y)\big\rangle_{\rho_i^{(\Phi)}}
\Big]
\nonumber\\
&\quad
+\frac12
\left(
-\frac{1}{2\hbar^2}\frac{\kappa^2}{4}
\right)^2
\int d^4x\,d^4y\,
\Big[
-4\,
T_{\Delta}^{\rm sys\,\mu\nu}(x)
\mathcal N_{\mu\nu\alpha\beta}(x,y)
T_{\Delta}^{\rm sys\,\alpha\beta}(y)
\nonumber\\
&\qquad
+4i\,
T_{\Delta}^{\rm sys\,\mu\nu}(x)
\mathcal D_{\mu\nu\alpha\beta}(x,y)
T_c^{\rm sys\,\alpha\beta}(y)
-4\,
T_{\Delta}^{\rm sys\,\mu\nu}(x)
\mathcal N^{\rm mix}_{\mu\nu\alpha\beta}(x,y)
T_{\Delta}^{\rm sys\,\alpha\beta}(y)
\Big],
\label{eq:SIF_core_complete}
\end{align}
where expectation values with respect to a density operator $\rho$ are defined by $\big\langle \hat{\mathcal O}\big\rangle_{\rho}:=\mathrm{Tr}\!\left(\rho\,\hat{\mathcal O}\right)$.

The intrinsic fluctuations of the metric perturbation are characterized by the graviton dissipation and noise kernels,
\begin{align}
\mathcal D^{(h)}_{\mu\nu\alpha\beta}(x,y)
&:=
i\,\theta(x^0-y^0)\,
\big\langle
[\hat h_{\mu\nu}(x),\hat h_{\alpha\beta}(y)]
\big\rangle_{\rho_i^{(h)}},
\label{eq:D_h_def}
\\
\mathcal N^{(h)}_{\mu\nu\alpha\beta}(x,y)
&:=
\frac12\,
\big\langle
\{\hat h_{\mu\nu}(x),\hat h_{\alpha\beta}(y)\}
\big\rangle_{\rho_i^{(h)}}.
\label{eq:N_h_def}
\end{align}
Here $\mathcal D^{(h)}$ is the retarded graviton response function
and $\mathcal N^{(h)}$ is the corresponding Hadamard correlator.

The scalar environment contributes through composite kernels
obtained by dressing the stress--energy tensor correlators of the
scalar field with graviton propagators,
\begin{align}
\mathcal D_{\mu\nu\alpha\beta}(x,y)
&=
\int d^4z\,d^4z'\,
G^{(h)R}{}_{\mu\nu}{}^{\rho\sigma}(x,z)\,
\mathcal D^{(\Phi)}_{\rho\sigma\lambda\gamma}(z,z')\,
G^{(h)A}{}^{\lambda\gamma}{}_{\alpha\beta}(z',y),
\label{eq:core_D_def}
\\
\mathcal N_{\mu\nu\alpha\beta}(x,y)
&=
\int d^4z\,d^4z'\,
G^{(h)R}{}_{\mu\nu}{}^{\rho\sigma}(x,z)\,
N^{(\Phi)}_{\rho\sigma\lambda\gamma}(z,z')\,
G^{(h)A}{}^{\lambda\gamma}{}_{\alpha\beta}(z',y),
\label{eq:core_N_def1}
\\
\mathcal N^{\rm mix}_{\mu\nu\alpha\beta}(x,y)
&=
\int d^4z\,d^4z'\,
G^{(h)R}{}_{\mu\nu}{}^{\rho\sigma}(x,z)\,
\mathcal D^{(\Phi)}_{\rho\sigma\lambda\gamma}(z,z')\,
\mathcal N^{(h)\,\lambda\gamma}{}_{\alpha\beta}(z',y),
\label{eq:core_N_def2}
\end{align}
where $G^{(h)R}$ and $G^{(h)A}$ denote the retarded and advanced graviton Green's functions, with $G^{(h)R}_{\mu\nu\alpha\beta}(x,y) \equiv \mathcal D^{(h)}_{\mu\nu\alpha\beta}(x,y)$ and $G^{(h)A}(x,y)=G^{(h)R}(y,x)$.
The scalar stress--energy tensor correlators are defined by
\begin{align}
\mathcal D^{(\Phi)}_{\rho\sigma\lambda\gamma}(z,z')
&:=
i\,\theta(z^0-z'^0)\,
\big\langle
[T^\Phi_{\rho\sigma}(z),T^\Phi_{\lambda\gamma}(z')]
\big\rangle_{\rho_i^{(\Phi)}, {\rm conn}},
\label{eq:D_Phi_def}
\\
N^{(\Phi)}_{\rho\sigma\lambda\gamma}(z,z')
&:=
\frac12\,
\big\langle
\{T^\Phi_{\rho\sigma}(z),T^\Phi_{\lambda\gamma}(z')\}
\big\rangle_{\rho_i^{(\Phi)}, {\rm conn}},
\label{eq:N_Phi_def}
\end{align}
where ``conn'' denotes connected parts.
Equation~\eqref{eq:SIF_core_complete} therefore provides the coarse-grained action governing the reduced dynamics of the system worldline.

\section{Decoherence functional}
\label{sec:decoherence_functional}
In this section we construct a modeled decoherence function suitable for physical evaluation in the nonrelativistic regime relevant to macroscopic experiments.
We begin with an idealized microscopic model in which the environment is described by finite-width wave packets, while the system is treated as a localized source.
The resulting decoherence functional is expressed in terms of a noise kernel and a system source, both defined as distributions.
Their direct pairing in the bilinear form is therefore not automatically well-defined and obscures the finite temporal and spatial scales relevant to realistic experiments.
We introduce a controlled modeling procedure based on coarse graining in space and time, implemented through windowing and convolution.
This procedure yields a well-defined bilinear functional that incorporates finite spacetime support and finite resolution scales.

\subsection{Decoherence functional}
\label{subsec:decoherence_functional}
The decoherence functional is defined as the real part of the influence action,
\begin{align}
\Gamma[X_+,X_-]
:=
-\,\Re\!\left(\frac{i}{\hbar}S_{\rm IF}[X_\pm]\right).
\label{eq:Gamma_def}
\end{align}
Keeping the dominant contribution from the effective noise kernel, which encodes stochastic fluctuations of the environment, the decoherence functional takes the form,
\begin{align}
\Gamma[X_+,X_-]
&=
\frac{\kappa^4}{32\hbar^4}
\int d^4x\,d^4y\;
T^{\rm sys}_{\Delta\,\mu\nu}(x)\,
\mathcal N^{\mu\nu\alpha\beta}(x,y)\,
T^{\rm sys}_{\Delta\,\alpha\beta}(y)\,.
\label{eq:Gamma_xy_formal}
\end{align}
Without assuming translation invariance, we introduce the double Fourier transform
\begin{align}
\widetilde{\mathcal N}_{\mu\nu\alpha\beta}(k,k')
:=
\int d^4x\,d^4y\;
e^{-ik\cdot x}\,
\mathcal N_{\mu\nu\alpha\beta}(x,y)\,
e^{+ik'\cdot y},
\label{eq:N_doubleFT_def_paper}
\end{align}
so that Eq.~\eqref{eq:Gamma_xy_formal} becomes
\begin{align}
\Gamma[X_+,X_-]
=
\frac{\kappa^4}{32\hbar^4}
\int \frac{d^4k}{(2\pi)^4}\frac{d^4k'}{(2\pi)^4}\;
\widetilde T^{\rm sys}_{\Delta\,\mu\nu}(-k)\,
\widetilde{\mathcal N}^{\mu\nu\alpha\beta}(k,k')\,
\widetilde T^{\rm sys}_{\Delta\,\alpha\beta}(k').
\label{eq:Gamma_double_formal}
\end{align}
Equation~\eqref{eq:Gamma_xy_formal} should be understood as the evaluation of the bilinear distribution on a pair of test functions, $\mathcal N[F,G] := \int d^4x\,d^4y\; F_{\mu\nu}(x)\,\mathcal N^{\mu\nu\alpha\beta}(x,y)\,G_{\alpha\beta}(y)$.
For an ideal point-particle superposition, however, $T^{\rm sys}_\Delta$ is distributional, so that the pairing $\mathcal N[T^{\rm sys}_\Delta,T^{\rm sys}_\Delta]$ is not well defined.

In this paper, we treat the microscopic model as an idealized description and incorporate experimental limitations through mathematical coarse graining of the resulting distributions.
This modeling procedure renders the decoherence functional well defined while effectively incorporating finite spacetime scales relevant to a realistic experiment and preserving the structure of the underlying microscopic dynamics.

Our modeling procedure uses two standard operations on distributions: windowing and convolution. 
Windowing localizes observables in spacetime, while convolution introduces a finite resolution.
For a one-point distribution $O(x)$ we define
\begin{align}
O^{(f)}(x) := f(x)\,O(x),\qquad 
O_{(f)}(x) := (f*O)(x),
\label{eq:model_1pt_paper}
\end{align}
where $f$ is a smooth profile.
In the same manner, for a two-point distribution $W(x,y)$ with formal kernel representation, the corresponding operations are defined by
\begin{align}
W^{(f,g)}(x,y)
:= f(x)\,W(x,y)\,g(y),\qquad
W_{(f,g)}(x,y)
:= (f*W*g)(x,y),
\label{eq:model_2pt_paper}
\end{align}
where $f, g$ are smooth profiles.

These operations are applied to the environmental kernel and the system source.
The resulting quantities are denoted by $\mathcal N_{\mathrm{model}}$ and $T_{\Delta,\mathrm{model}}$, whose explicit forms are given later in Eqs.~\eqref{eq:modeled_N_Phi} and \eqref{eq:T00Delta_model_paper}.
The modeled decoherence functional becomes
\begin{align}
\Gamma[X_+,X_-]
&=
\frac{\kappa^4}{32\hbar^4}
\int\frac{d^4k}{(2\pi)^4}\frac{d^4k'}{(2\pi)^4}\;
\widetilde{T}_{\Delta,\mathrm{model}}{}^{\mu\nu}(-k)\,
\widetilde{\mathcal N}_{\mathrm{model}\,\mu\nu\alpha\beta}(k,k')\,
\widetilde{T}_{\Delta,\mathrm{model}}{}^{\alpha\beta}(k').
\label{eq:Gamma_model_def}
\end{align}
Because spacetime windowing breaks translation invariance, the modeled kernel depends on two independent four-momenta $k$ and $k'$.
In the following we restrict our attention to the $00$ component of the stress--energy tensor, which dominates in the nonrelativistic regime.

\subsection{Environmental noise kernel}
\label{subsec:noise_model}

In this subsection we construct the modeled environmental noise kernel in the nonrelativistic dilute regime, based on an idealized wave-packet description of the environmental gas particles.

Decomposing the complex scalar field as
\begin{align}
\Phi(x)
=
\frac{e^{-imt}}{\sqrt{2m}}\psi(x)
+
\frac{e^{+imt}}{\sqrt{2m}}\chi^\dagger(x),
\label{eq:Phi_NR_split}
\end{align}
and neglecting the antiparticle sector $\chi$, we obtain to leading order in $|\bm p|/m$
\begin{align}
T^{00}(x)\simeq m\,\hat n(x),
\qquad
\hat n(x):=\psi^\dagger(x)\psi(x).
\label{eq:T00_NR_density}
\end{align}
Accordingly, the $0000$ component of the scalar noise kernel~\eqref{eq:N_Phi_def} reduces to the density-density correlator,
\begin{align}
N^{(\Phi)}_{0000}(x,y)
\simeq
m^2 N_{nn}(x,y),
\qquad
N_{nn}(x,y):=
\frac12\big\langle\{\hat n(x),\hat n(y)\}\big\rangle_{\rm conn}.
\label{eq:N_phi_density}
\end{align}
Using the standard plane-wave quantization of $\psi$ with annihilation operators $\hat a_{\bm p}$ satisfying $[\hat a_{\bm p},\hat a^\dagger_{\bm q}] =(2\pi)^3\delta^{(3)}(\bm p-\bm q)$, the Fourier mode of the density operator becomes
\begin{align}
\hat n(k)
=
2\pi\int\frac{d^3p}{(2\pi)^3}\;
\hat a^\dagger_{\bm p}\hat a_{\bm p+\bm k}\,
\delta\Big(\omega-\big(\varepsilon_{\bm p+\bm k}-\varepsilon_{\bm p}\big)\Big)\,,
\label{eq:nk_w_aadag}
\end{align}
where $\varepsilon_{\bm p}=\bm p^2/(2m)$ gives the nonrelativistic single-particle energy.

The next step is to specify the state of the gas.
Since we are interested in a dilute Maxwell--Boltzmann regime, it is natural to describe the environment by its one-body reduced density matrix.
We introduce a Gaussian wave-packet state centered at position $\bm X$,
\begin{align}
|\psi_{v,\Omega_{\rm in};\bm X}\rangle
=
\int \frac{d^3p}{(2\pi)^3}\;
e^{-i\bm p\cdot\bm X}\,
\phi(\bm p;v,\Omega_{\rm in},\sigma_k)\,
|\bm p\rangle,
\end{align}
where the momentum-space wavefunction is 
\begin{align}
\phi(\bm p;v,\Omega_{\rm in},\sigma_k)
=
(2\pi)^{3/2}
(2\pi\sigma_k^2)^{-3/4}
\exp\!\left[
-\frac{ |\bm p-p_0\bm\Omega_{\rm in}|^2}{4\sigma_k^2}
\right],
\end{align}
with $p_0 := mv$ the central momentum of the wave packet directed along
$\bm\Omega_{\rm in}$, and $\sigma_k$ the momentum-space  width of the packet.
We model the normalized one-body reduced density matrix as an incoherent mixture of such wave packets with speed distribution $\mathcal P(v)$, isotropic incoming directions $\bm \Omega_{\rm in}$, and a spatial profile $u(\bm X)$ describing the distribution of packet centers,
\begin{align}
\rho_1
=
\int_0^\infty dv\,\mathcal P(v)
\int\frac{d\Omega_{\rm in}}{4\pi}
\int d^3X\,u(\bm X)\;
|\psi_{v,\Omega_{\rm in};\bm X}\rangle
\langle\psi_{v,\Omega_{\rm in};\bm X}|.
\label{eq:rho1_model_paper}
\end{align}

For concreteness, we now specialize to a homogeneous gas and take
\begin{align}
u(\bm X)=V^{-1}\mathbf 1_V(\bm X),
\qquad
\tilde u(\bm K)\simeq\frac{(2\pi)^3}{V}\delta^{(3)}(\bm K),
\label{eq:homogeneous_u_here}
\end{align}
where $V$ is a large normalization volume used to define the homogeneous limit.
At this stage we adopt this idealized description, deferring the implementation of finite resolution to the coarse-graining procedure introduced below.

The equal-time two-point function is then parametrized by
\begin{align}
\langle a^\dagger_{\bm p_1}a_{\bm p_2}\rangle
=
N_{\rm eff}\,\langle \bm p_2|\rho_1|\bm p_1\rangle,
\label{eq:adag_a_rho1_paper}
\end{align}
where $\bm p_1$ and $\bm p_2$ are generic single-particle plane-wave momenta, and $N_{\rm eff}$ denotes the effective number of particles contained in the normalization volume $V$.
We define the corresponding homogeneous particle density by $n:={N_{\rm eff}}/{V}$.
Evaluating \eqref{eq:adag_a_rho1_paper} with the Gaussian wave-packet ensemble gives
\begin{align}
\langle a^\dagger_{\bm p_1} a_{\bm p_2}\rangle
&=
(2\pi)^3N_{\rm eff}\,\tilde u(\bm p_2-\bm p_1)
\int_0^\infty dv\,\mathcal P(v)\,
(2\pi\sigma_k^2)^{-3/2}
\exp\!\left[
-\frac{p_1^2+p_2^2+2p_0^2}{4\sigma_k^2}
\right]
\frac{\sinh\!\left(\frac{p_0|\bm p_1+\bm p_2|}{2\sigma_k^2}\right)}
     {\frac{p_0|\bm p_1+\bm p_2|}{2\sigma_k^2}}.
\label{eq:adag_a_wavepacket_paper}
\end{align}

In the idealized description above, the environmental correlations are defined over an infinite and homogeneous spacetime region.
In a realistic experiment, however, environmental fluctuations are sampled only within a finite spacetime domain.
To model this finite spacetime support, we introduce spacetime windowing of the environmental kernel.
We multiply the two-point distribution by a factorized window
\begin{align}
f(x):=\chi_T(t)\,\varphi_r(\bm x),
\label{eq:f_window_def}
\end{align}
where $\chi_T$ and $\varphi_r$ characterize the temporal and spatial support of the sampled correlations, with $T$ and $r$ denoting the associated coarse-graining scales.
Their explicit forms and interpretations depend on the experimental and analytical settings of interest and will be specified later.
The corresponding modeled kernel is defined by
\begin{align}N^{(\Phi)}_{{\rm model}\,\mu\nu\alpha\beta}(x,y)
:=
\left(N^{(\Phi)}_{\mu\nu\alpha\beta}\right)^{(f,f)}(x,y)\,.
\label{eq:modeled_N_Phi}
\end{align}

\begin{figure}[t]
\centering
\begin{tikzpicture}[
  xscale=1.08,
  yscale=1.0,
  graviton/.style={
    decorate,
    decoration={snake, amplitude=1.1mm, segment length=4mm},
    thick
  },
  solid/.style={thick},
  flow/.style={thick, -{Latex[length=2mm]}},
  dflow/.style={thick, dashed},
  dot/.style={circle, fill=black, inner sep=1.7pt},
  lab/.style={font=\small},
  mom/.style={font=\small}
]

\coordinate (A) at (-4.6,0);
\coordinate (B) at (-3.0,0);
\coordinate (C) at (-1.05,0);
\coordinate (D) at ( 1.05,0);
\coordinate (E) at ( 3.0,0);
\coordinate (F) at ( 4.6,0);

\draw[graviton] (A) -- (B);
\draw[solid]    (B) -- (C);
\draw[solid]    (D) -- (E);
\draw[graviton] (E) -- (F);

\node[dot] at (B) {};
\node[dot] at (C) {};
\node[dot] at (D) {};
\node[dot] at (E) {};

\draw[dflow]
  (C) arc[start angle=180,end angle=0,x radius=1.05,y radius=0.85];

\draw[flow]
  (D) arc[start angle=0,end angle=-180,x radius=1.05,y radius=0.85];

\node[lab, above left=1pt and 1pt of C] {$n(q)$};
\node[lab, above right=1pt and 1pt of D] {$n(- q')$};

\node[lab] at (-3.85,0.55) {$(\omega,\bm k)$};
\node[lab] at ( 3.85,0.55) {$(\omega',\bm k')$};

\node[lab] at (0,1.20)
  {$\langle a_{\bm p+\bm q}a^\dagger_{\bm p'}\rangle
  \simeq (2\pi)^3\delta^{(3)}(\bm p+\bm q-\bm p')$};

\node[lab] at (0,-1.20)
  {$\langle a^\dagger_{\bm p}a_{\bm p'-\bm q'}\rangle
  \;\to\;
  \langle a^\dagger_{\bm p}a_{\bm p+\bm q-\bm q'}\rangle$};

\node[lab, below] at (-3.05,-0.38)
  {$\widetilde\chi_T(\omega-\omega_{\bm q})
    \widetilde\varphi_r(\bm k-\bm q)$};

\node[lab, below] at (3.05,-0.38)
  {$\widetilde\chi_T(\omega_{\bm q'}-\omega')
    \widetilde\varphi_r(\bm q'-\bm k')$};

\end{tikzpicture}
\caption{
Diagrammatic representation of the Wick contraction contributing to the windowed density-density correlator entering $\widetilde N^{(\Phi)}_{{\rm model},0000}(k,k')$.
The central vertices denote the density insertions
$n(q)$ and $n(-q')$.
The upper contraction reduces, in the dilute Maxwell--Boltzmann limit,
to $(2\pi)^3\delta^{(3)}(\bm p+\bm q-\bm p')$,
thereby fixing the intermediate momentum $\bm p'$.
The lower contraction yields the remaining one-body density matrix $\langle a^\dagger_{\bm p}a_{\bm p+\bm q-\bm q'}\rangle = \langle a^\dagger_{\bm P - \bm Q/2}a_{\bm P + \bm Q/2} \rangle$.
The external momenta $(\omega,\bm k)$ and $(\omega',\bm k')$ are connected to the density insertions through the temporal and spatial window functions $\widetilde\chi_T$ and $\widetilde\varphi_r$.
}
\label{fig:windowed_density_contraction}
\end{figure}
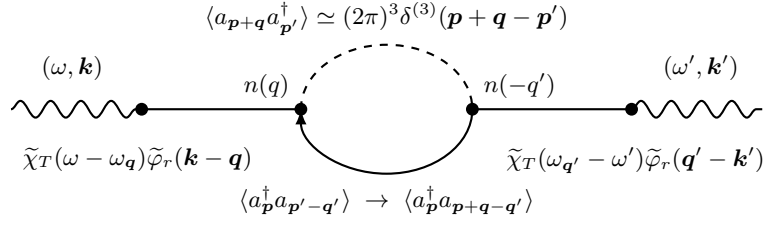

Applying this operation to the density-density kernel~\eqref{eq:N_phi_density}, substituting Eq.~\eqref{eq:nk_w_aadag}, and carrying out the Wick contraction with Eq.~\eqref{eq:adag_a_wavepacket_paper}, the spatial windowing generates convolutions between the kernel momenta $k=(\omega,\bm k)$ and $k'=(\omega',\bm k')$ and the momentum transfers entering the two density operators, denoted by $\bm q$ and $\bm q'$. 
These variables are related through the window factors $\tilde\varphi_r(\bm k-\bm q)$ and $\tilde\varphi_r(\bm q'-\bm k')$.
In the absence of windowing, the convolution reduces to $\bm q=\bm k$ and $\bm q'=\bm k'$.
The structure of the Wick contraction and the associated momentum flow are illustrated schematically in Fig.~\ref{fig:windowed_density_contraction}.

The variable $\bm p$ in Eq.~\eqref{eq:nk_w_aadag} is the single-particle plane-wave momentum integrated over in the density operator.
After the Wick contraction, the second particle momentum is fixed, and the remaining one-body density matrix takes the form $\braket{a^\dagger_{\bm p}a_{\bm p+\bm Q}}$ with $\bm Q:=\bm q-\bm q'$. 
Introducing the symmetrized momentum $\bm P:=\bm  p+(\bm Q/{2})$, this matrix element is written as $\braket{a^\dagger_{\bm P-\bm Q/2}a_{\bm P+\bm Q/2}}$.
Thus, $\bm Q$ is the momentum separation in the one-body density matrix, while $\bm P$ is the corresponding average single-particle momentum.

With these conventions, the modeled matter noise kernel becomes
\begin{align}
\widetilde{ N}^{(\Phi)}_{{\rm model}, \,0000}(k,k')
&\simeq
\frac{m^2}{2(2\pi)^3}
\int\frac{d^3P}{(2\pi)^3}
\int d^3q\,d^3q'\;
\tilde\chi_T\!\left(\omega-\omega_{\bm q}(\bm P,\bm q')\right)\,
\tilde\chi_T\!\left(\omega_{\bm q'}(\bm P,\bm q)-\omega'\right)
\nonumber\\
&\qquad\times
\tilde \varphi_r(\bm k-\bm q)\,\tilde \varphi_r(\bm q'-\bm k')\,
(2\pi)^3N_{\rm eff}\,\tilde u(\bm Q)\nonumber\\
&\qquad\times\int_0^\infty dv\,\mathcal P(v)\,(2\pi\sigma_k^2)^{-3/2}
\exp\!\left[
-\frac{\bm P^2+p_0^2}{2\sigma_k^2}
-\frac{\bm Q^2}{8\sigma_k^2}
\right]
\frac{\sinh\!\left(\frac{p_0|\bm P|}{\sigma_k^2}\right)}
     {\frac{p_0|\bm P|}{\sigma_k^2}}
\nonumber\\
&\qquad
+\Big((\omega,\omega',\bm q,\bm q')\rightarrow(-\omega,-\omega',-\bm q,-\bm q')\Big),
\label{eq:N0000_model_with_wavepacket}
\end{align}
where $p_0:=mv$ denotes the central momentum of each Gaussian wave packet, and
\begin{align}
    \omega_{\bm q}(\bm P,\bm q')  := \epsilon_{\bm p+\bm q}-\epsilon_{\bm p} = \frac{(\bm P+\bm q'/2)\cdot\bm q}{m}\,,\qquad
    \omega_{\bm q'}(\bm P,\bm q)  := \epsilon_{\bm p+\bm q}-\epsilon_{\bm p + \bm q -\bm q'} = \frac{(\bm P+\bm q/2)\cdot\bm q'}{m}.
    \label{eq:omega_q_qprime_def}
\end{align}
The dependence on $\chi_T$ and $\varphi_r$ encodes the finite spacetime region over which the environmental correlations are sampled. 
The dependence on $\sigma_k$, $\mathcal P(v)$, and $\tilde{u}$ encodes the underlying state of the idealized environment.

The effective modeled kernel entering the decoherence functional is obtained by dressing the kernel with retarded and advanced graviton propagators,
\begin{align}
\widetilde{\mathcal N}_{{\rm model}\, \mu\nu\alpha\beta}(k,k')
=
G^{(h)R}_{\mu\nu}{}^{\rho\sigma}(k)\,
\widetilde{ N}^{(\Phi)}_{{\rm model} \,\rho\sigma\lambda\gamma}(k,k')\,
G^{(h)A\,\lambda\gamma}{}_{\alpha\beta}(k').
\label{eq:N_model_dressed_def}
\end{align}
In the nonrelativistic evaluation below, only the $0000$ component will be retained.

\subsection{Modeled system difference source}
\label{subsec:TDelta_model}

The formal system source appearing in the influence action is the difference between two point-particle stress--energy tensors supported on the forward and backward branches.
However, this source is singular and carries no information about the finite size of the system.
We therefore model it by spatial convolution with a normalized profile $w_{R_s}(\bm x)$, where $R_s$ denotes the effective size of the system.

In the quasi-static regime we approximate the two branches by
\begin{align}
\bm X_\pm(t)\simeq \bm X_c \pm \frac{\Delta\bm x}{2},
\label{eq:Xpm_static_def}
\end{align}
where $\bm X_c$ denotes the center of the superposition and $\Delta\bm x$ is the branch separation.
The ideal point-particle difference source is then
\begin{align}
T^{{\rm sys}\,00}_{\Delta}(t,\bm x)
=
M\Big[\delta^{(3)}(\bm x-\bm X_+)-\delta^{(3)}(\bm x-\bm X_-)\Big].
\label{eq:T00Delta_pt_paper}
\end{align}
Note that, for a physical mass distribution obeying momentum conservation, the leading nontrivial contribution begins at the quadrupole order.
In a realistic experimental setup, the effective dipole-like displacement of the system must be accompanied by compensating mass rearrangements in the surrounding apparatus.
Environmental particles at distances larger than the characteristic size of this system--apparatus configuration then probe the combined multipole structure, for which the net dipole moment is expected to cancel and the leading contribution crosses over to a more rapidly decaying higher-multipole behavior.
Here, we instead adopt the simplified dipole-like parametrization~\eqref{eq:Xpm_static_def} as an effective description of the branch separation, which provides a conservative estimate (i.e. an overestimate) of the decoherence effect.

We define the modeled source by convolution in the spatial coordinates,
\begin{align}
T^{00}_{\Delta,{\rm model}}(t,\bm x)
:=
\Big(T^{{\rm sys}\,00}_{\Delta}\Big)_{(w_{R_s})}(t,\bm x)
=
\int d^3y\;w_{R_s}(\bm x-\bm y)\,T^{{\rm sys}\,00}_{\Delta}(t,\bm y),
\label{eq:T00Delta_model_paper}
\end{align}
with normalization $\int d^3x\;w_{R_s}(\bm x)=1$.
This gives
\begin{align}
T^{00}_{\Delta,{\rm model}}(t,\bm x)
=
M\Big[w_{R_s}(\bm x-\bm X_+)-w_{R_s}(\bm x-\bm X_-)\Big].
\label{eq:T00Delta_model_explicit}
\end{align}
Taking the Fourier transform, we find
\begin{align}
\widetilde T^{00}_{\Delta,{\rm model}}(\omega,\bm k)
&=
-2iM\,\tilde w_{R_s}(\bm k)\,
e^{-i\bm k\cdot\bm X_c}\,
\sin\!\Big(\frac{\bm k\cdot\Delta\bm x}{2}\Big)\,
(2\pi)\delta(\omega).
\label{eq:Ttilde_smeared}
\end{align}
This formula isolates the three ingredients that later control the decoherence integral: the finite source profile $\tilde w_{R_s}$, the center position $\bm X_c$, and the branch separation $\Delta\bm x$.

\subsection{Nonrelativistic decoherence function}
\label{subsec:Gamma_NR}

We now combine the modeled environmental kernel and the modeled source.
In the quasi-static nonrelativistic limit the gravitational dressing is dominated by the $0000$ component, for which
\begin{align}
G^{(h)R}_{0000}(0,\bm k)=-\frac{\hbar}{2|\bm k|^2},
\qquad
G^{(h)A}_{0000}(0,\bm k)=\Big(G^{(h)R}_{0000}(0,\bm k)\Big)^*.
\label{eq:GR_GA_static_0000}
\end{align}
Retaining only the leading $00$ contribution, and noting that the term obtained from $(\bm q,\bm q')\rightarrow(-\bm q,-\bm q')$ gives the same contribution after integration over $\bm q$ and $\bm q'$, we substitute Eq.~\eqref{eq:N_model_dressed_def} together with Eqs.~\eqref{eq:N0000_model_with_wavepacket} and \eqref{eq:Ttilde_smeared} into \eqref{eq:Gamma_model_def}. Performing the $\omega$ and $\omega'$ integrals, we obtain
\begin{align}
\Gamma
&\simeq
\frac{\kappa^4}{32\hbar^2}\,M^2\,
m^2N_{\rm eff}
\int\frac{d^3k}{(2\pi)^3}\frac{d^3k'}{(2\pi)^3}\;
\frac{1}{k^2k'^2}\,
\tilde w_{R_s}(-\bm k)\,\tilde w_{R_s}(\bm k')\,
e^{i(\bm k-\bm k')\cdot\bm X_c}\,
\sin\!\Big(\frac{\bm k\cdot\Delta\bm x}{2}\Big)\,
\sin\!\Big(\frac{\bm k'\cdot\Delta\bm x}{2}\Big)
\nonumber\\
&\quad\times
\int\frac{d^3P}{(2\pi)^3}
\int d^3q\,d^3q'\;
\tilde\chi_T\!\left(-\frac{(\bm P+\bm q'/2)\cdot\bm q}{m}\right)\,
\tilde\chi_T\!\left(\frac{(\bm P+\bm q/2)\cdot\bm q'}{m}\right)
\tilde \varphi_r(\bm k-\bm q)\,\tilde \varphi_r(\bm q'-\bm k')
\nonumber\\
&\quad\times
\tilde u(\bm q-\bm q')
\int_0^\infty dv\,\mathcal P(v)\,(2\pi\sigma_k^2)^{-3/2}
\exp\!\left[
-\frac{\bm P^2+p_0^2}{2\sigma_k^2}
-\frac{(\bm q-\bm q')^2}{8\sigma_k^2}
\right]
\frac{\sinh\!\left(\frac{p_0|\bm P|}{\sigma_k^2}\right)}
     {\frac{p_0|\bm P|}{\sigma_k^2}}.
\label{eq:Gamma_full_wavepacket}
\end{align}
Equation~\eqref{eq:Gamma_full_wavepacket} is the starting point for the explicit evaluation carried out in the next section and provides a bridge between the formal CTP kernel and the experimentally relevant decoherence function.

\section{Evaluation of the decoherence function}
\label{sec:decoherence_evaluation}

In this section we evaluate the modeled decoherence function in a concrete class of nonrelativistic models.
We specialize to a homogeneous dilute gas described by a thermal Maxwell--Boltzmann velocity distribution, together with a Gaussian temporal window, a finite spherical spatial window, and a Gaussian finite-size profile for the system source.
The spherical window defines a cumulative decoherence function, denoted by $\Gamma_{<r}$, which represents the contribution obtained by sampling environmental fluctuations inside a sphere of radius $r$ centered at the system.
We analyze how the contribution builds up as the spatial window is enlarged, identify the momentum region responsible for the dominant large-radius contribution, and separately determine the behavior in the small-window regime.
The saturated decoherence function is then obtained by taking the large-radius limit of the resulting cumulative function.
In this section we set $\hbar=1$ unless explicitly restored.

\subsection{Homogeneous gas and spherical spatial window}
\label{subsec:decoherence_spherical}

We now specialize the nonrelativistic decoherence function~\eqref{eq:Gamma_full_wavepacket} to a homogeneous dilute gas characterized by a Maxwell--Boltzmann velocity distribution at temperature $T_{\rm gas}$.
The velocity distribution is taken to be the Maxwell--Boltzmann form
\begin{align}
\mathcal P(v)
=
4\pi v^2
\left(\frac{m}{2\pi k_B T_{\rm gas}}\right)^{3/2}
\exp\!\left(-\frac{m v^2}{2k_B T_{\rm gas}}\right).
\label{eq:MB_distribution}
\end{align}
We further set the momentum-space width of the environmental wave packets to the rms thermal momentum scale,
\begin{align}
\sigma_k
=
\sqrt{
3mk_B T_{\rm gas}
},
\label{eq:sigma_k_thermal}
\end{align}
and identify the homogeneous particle density as
\begin{align}
n
=
\frac{P_{\rm gas}}{k_B T_{\rm gas}}.
\label{eq:n_thermal}
\end{align}
The corresponding typical thermal velocity is given by
\begin{align}
v_{\rm typ}
:=
\sqrt{
\frac{8k_B T_{\rm gas}}{\pi m}
}.
\label{eq:v_typ_MB}
\end{align}

We next specify the spacetime windows.
To keep the analysis as independent of a particular experimental setup as possible, we characterize the coarse graining by common temporal and spatial scales.
The window functions $\chi_T$ and $\varphi_r$ effectively model finite temporal observation and finite spatial extent of the environmental gas, respectively.
We require that the window functions approach unity in the large-scale limit, $\lim_{T\to\infty}\chi_T(t)=1$ and $\lim_{r\to\infty}\varphi_r(\bm x)=1$, so that the original kernel is recovered.
For the temporal window we choose a Gaussian profile
\begin{align}
\chi_T(t)
=
\exp\!\left(-\frac{t^2}{2T^2}\right),
\qquad
\tilde\chi_T(\omega)
=
\sqrt{2\pi}\,T\,e^{-T^2\omega^2/2},
\label{eq:Gaussian_time_window}
\end{align}
where $T$ characterizes the observation time.
For the spatial window we use a spherical profile
\begin{align}
\varphi_r(\bm x)
=
\Theta(r-|\bm x|),
\label{eq:spherical_window_phi}
\end{align}
where $r$ characterizes the effective spatial extent of the environmental gas incorporated through the coarse graining.
This $\varphi_r$ may be regarded as the sharp-edge limit of a smooth spherical profile.
Its Fourier transform is
\begin{align}
\tilde\varphi_r(\bm Q)
=
\int_{|\bm x|<r}d^3x\,e^{-i\bm Q\cdot\bm x}
=
V_r F(Qr),
\qquad
V_r=\frac{4\pi r^3}{3},
\label{eq:spherical_window_FT_1}
\end{align}
with
\begin{align}
F(x)
=
3\,\frac{\sin x-x\cos x}{x^3}.
\label{eq:spherical_window_FT_2}
\end{align}

In the following we center the spatial window on the superposition center of the system, which amounts to setting $\bm X_c=0$.
For the finite-size profile of the system source we use the normalized Gaussian
\begin{align}
w_{R_s}(\bm x)
=
\frac{1}{(2\pi R_s^2)^{3/2}}
\exp\!\left(-\frac{\bm x^2}{2R_s^2}\right),
\qquad
\tilde w_{R_s}(\bm k)
=
\exp\!\left(-\frac{R_s^2\bm k^2}{2}\right),
\label{eq:wR_gaussian_FT}
\end{align}
where $R_s$ characterizes the physical size of the system source.

We denote by $\Gamma_{<r}$ the decoherence function obtained from Eq.~\eqref{eq:Gamma_full_wavepacket} with these windows.
The large-radius limit of the cumulative decoherence function gives the total contribution of the homogeneous gas,
\begin{align}
\Gamma
=
\lim_{r\to\infty}\Gamma_{<r}.
\end{align}

Using the relation~\eqref{eq:homogeneous_u_here}, the $(\bm q,\bm q')$ sector in Eq.~\eqref{eq:Gamma_full_wavepacket} reduces to a single integral over the momentum transfer $\bm q$.
The cumulative decoherence function can then be written as
\begin{align}
\Gamma_{<r}
&\simeq
\frac{\kappa^4}{32}M^2m^2 n
\int\frac{d^3k}{(2\pi)^3}
\frac{d^3k'}{(2\pi)^3}
\frac{1}{k^2k'^2}\,
\tilde w_{R_s}(-\bm k)\tilde w_{R_s}(\bm k')
\sin\!\Big(\frac{\bm k\cdot\Delta\bm x}{2}\Big)
\sin\!\Big(\frac{\bm k'\cdot\Delta\bm x}{2}\Big)
\,I_r(\bm k,\bm k'),
\label{eq:Gamma_window_spatial}
\end{align}
with
\begin{align}
I_r(\bm k,\bm k')
:=
(2\pi)^3
\int d^3q\;
\tilde\varphi_r(\bm k-\bm q)\,
\tilde\varphi_r(\bm q-\bm k')\,
\mathcal J(q).
\label{eq:Ir_window_def}
\end{align}
The rotationally invariant kernel $\mathcal J(q)$ is given by
\begin{align}
\mathcal J(q)
&:=
\int\frac{d^3P}{(2\pi)^3}
\left|
\tilde\chi_T\!\left(
\frac{(\bm P+\bm q/2)\cdot\bm q}{m}
\right)
\right|^2
\int_0^\infty dv\,\mathcal P(v)\,
(2\pi\sigma_k^2)^{-3/2}
\exp\!\left[
-\frac{\bm P^2+p_0^2}{2\sigma_k^2}
\right]
\frac{\sinh\!\left(
\frac{p_0|\bm P|}{\sigma_k^2}
\right)}
{\frac{p_0|\bm P|}{\sigma_k^2}} .
\label{eq:J_q_integral_def}
\end{align}

We next evaluate the $\bm P$ integral.
Choosing the polar axis along $\bm q$ and writing
$\bm P\cdot\bm q=Pq\mu$, we obtain
\begin{align}
\int d\Omega_{\bm P}\;
\Big|
\tilde\chi_T\!\Big(
\frac{\bm P\cdot\bm q}{m}
+
\frac{q^2}{2m}
\Big)
\Big|^2
&=
4\pi^2\sqrt{\pi}\,T
\frac{m}{Pq}
\frac{1}{2}
\left[
{\rm erf}\!\left(
\frac{Tq}{m}\left(P+\frac{q}{2}\right)
\right)
+
{\rm erf}\!\left(
\frac{Tq}{m}\left(P-\frac{q}{2}\right)
\right)
\right],
\label{eq:P_angular_integral}
\end{align}
where ${\rm erf}(z)$ is the error function defined by ${\rm erf}(z):=({2}/{\sqrt{\pi}})\int_0^z e^{-t^2}dt$.
The remaining radial integral can be evaluated using the shifted
Gaussian--error-function identity
\begin{align}
\int_{-\infty}^{\infty}
\frac{dP}{\sqrt{2\pi}\sigma}\,
e^{-\frac{(P-\mu)^2}{2\sigma^2}}\,
{\rm erf}\!\big(\gamma(P+a)\big)
=
{\rm erf}\!\left(
\frac{\gamma(\mu+a)}
{\sqrt{1+2\gamma^2\sigma^2}}
\right),
\label{eq:Gaussian_erf_identity}
\end{align}
which follows by differentiating the left-hand side with respect to $\mu$, using integration by parts and Gaussian convolution with the derivative of the error function, and then fixing the integration constant from the condition that the integral vanishes at $\mu=-a$.
Applying Eq.~\eqref{eq:Gaussian_erf_identity} with $\gamma={Tq}/{m},\ \mu=p_0,\ \sigma=\sigma_k,\ a=\pm q/2$, we obtain
\begin{align}
\mathcal J(q)
&=
\frac{T}{8\pi^{3/2}}
\int_0^\infty dv\,\mathcal P(v)\,
\frac{1}{v}\,
\frac{1}{q}\,
\frac{1}{2}
\left[
{\rm erf}\!\left(
\frac{Tq\left(v+\frac{q}{2m}\right)}
{\sqrt{1+(q/k_\sigma)^2}}
\right)
+
{\rm erf}\!\left(
\frac{Tq\left(v-\frac{q}{2m}\right)}
{\sqrt{1+(q/k_\sigma)^2}}
\right)
\right],
\label{eq:J_q_def}
\end{align}
where $k_\sigma:=m/(\sqrt{2}\,T\sigma_k)$ is the scale associated with the momentum width of the environmental wave packet.

For a fixed radius $r$, the spherical window introduces the momentum scale $r^{-1}$.
When
\begin{align}
kr,\ k'r \gtrsim 1,
\end{align}
the convolution in Eq.~\eqref{eq:Ir_window_def} becomes approximately diagonal $\bm q\simeq\bm k\simeq\bm k'$.
We therefore decompose the cumulative decoherence function into diagonal and off-diagonal parts,
\begin{align}
\Gamma_{<r}
=
\Gamma^{\rm diag}_{<r}
+
\Gamma^{\rm off}_{<r}.
\end{align}
The diagonal part is evaluated in the asymptotic regime
\begin{align}
\Gamma^{\rm diag}_{<r}
&:=
\frac{\kappa^4}{32}M^2m^2 n
\int_{k,k'\gtrsim r^{-1}}
\frac{d^3k}{(2\pi)^3}
\frac{d^3k'}{(2\pi)^3}
\frac{1}{k^2k'^2}\,
\tilde w_{R_s}(-\bm k)\tilde w_{R_s}(\bm k')
\nonumber\\
&\quad\times
\sin\!\Big(\frac{\bm k\cdot\Delta\bm x}{2}\Big)
\sin\!\Big(\frac{\bm k'\cdot\Delta\bm x}{2}\Big)
\,I_r(\bm k,\bm k'),
\end{align}
while the off-diagonal contribution is defined by
\begin{align}
\Gamma^{\rm off}_{<r}
:=
\Gamma_{<r}
-
\Gamma^{\rm diag}_{<r}.
\label{eq:Gamma_off_def}
\end{align}

In the following subsections, we evaluate these two contributions separately.
For sufficiently large $r$ compared with the spatial scales of the system, $R_s$ and $\Delta x$, the radius may be interpreted as an effective spatial extent of the environmental gas.
For smaller $r$, however, the result should instead be viewed as the small-window contribution obtained from spherical coarse graining of the idealized homogeneous kernel.
This regime nevertheless describes how the cumulative decoherence function begins to build up at small radii.

\subsection{Diagonal sector}
\label{subsec:decoherence_reduction}

We first evaluate the diagonal contribution $\Gamma^{\rm diag}_{<r}$.
In this sector, the width of $\tilde\varphi_r$ is small compared with the kernel momenta $k, k'$.
Over the support of the convolution in Eq.~\eqref{eq:Ir_window_def}, the momentum transfer is localized near $\bm q\simeq\bm k\simeq\bm k'$, which is justified by
$ |\bm q-\bm k|/k = O((kr)^{-1}) \ll 1$.
To leading order,
\begin{align}
I_r(\bm k,\bm k')
&\simeq
(2\pi)^3
\mathcal J(k)
\int d^3q\;
\tilde\varphi_r(\bm k-\bm q)\,
\tilde\varphi_r(\bm q-\bm k')
\nonumber\\
&=
(2\pi)^6
\widetilde{\varphi_r^2}(\bm k-\bm k')\,
\mathcal J(k).
\label{eq:Ir_diag_general}
\end{align}
Since the spherical window satisfies $\varphi_r^2=\varphi_r$, this becomes
\begin{align}
I_r(\bm k,\bm k')
\simeq
(2\pi)^6
\tilde\varphi_r(\bm k-\bm k')\,
\mathcal J(k).
\label{eq:Ir_window_diagonal}
\end{align}
Moreover, since
\begin{align}
\int\frac{d^3Q}{(2\pi)^3}
\tilde\varphi_r(\bm Q)
=
\varphi_r(\bm 0)=1,
\end{align}
and $\tilde\varphi_r(\bm k-\bm k')$ has width of order $r^{-1}$ in momentum space, the $\bm k'$ integral is localized near $\bm k'\simeq\bm k$.
Performing the angular integration over $\bm k$,
$\int d\Omega_{\bm k}\;
\sin^2\!\Big({\bm k\cdot\Delta\bm x}/{2}\Big)
=
2\pi\left[1-{\sin(k\Delta x)}/({k\Delta x})\right]$
and using $|\tilde w_{R_s}(\bm k)|^2 = e^{-R_s^2k^2}$,
the diagonal contribution reduces to the one-dimensional integral form
\begin{align}
\Gamma^{\rm diag}_{<r}
&\simeq
\mathcal C
\int_{r^{-1}}^\infty dk\;
\frac{e^{-R_s^2k^2}}{k^2}
\left(
1-\frac{\sin(k\Delta x)}{k\Delta x}
\right)
\mathcal J(k),
\label{eq:Gamma_1D_diag_r}
\end{align}
where
\begin{align}
\mathcal C
:=
\frac{(2\pi)^4\kappa^4}{32}
M^2m^2 n.
\label{eq:C_prefactor}
\end{align}

The behavior of the diagonal contribution~\eqref{eq:Gamma_1D_diag_r} is governed by the competition between several characteristic momentum scales.
The Gaussian source profile introduces an ultraviolet cutoff at
$k_{R_s}:=1/R_s$,
while the interference factor $1-\sin(k\Delta x)/(k\Delta x)$ defines the scale $k_\Delta:=1/\Delta x$.
The ultraviolet cutoff scale is therefore
\begin{align}
k_{\rm UV}
:=
\min(k_{R_s},k_\Delta).
\label{eq:kUV_def}
\end{align}

The kernel $\mathcal J(k)$ defined in Eq.~\eqref{eq:J_q_def} contains an additional scale associated with recoil.
The shift $k/(2m)$ becomes comparable to the velocity when $k\sim 2mv$, which defines the recoil scale $k_*:=2m v_{\rm typ}$.
Since the dominant contribution to Eq.~\eqref{eq:Gamma_1D_diag_r} comes from $k\lesssim k_{\rm UV}$, the recoil correction is negligible over the relevant integration region provided the recoil-free condition $k_{\rm UV}\ll k_*$ is satisfied.
Under the recoil-free condition, Eq.~\eqref{eq:J_q_def} reduces to
\begin{align}
\mathcal J(k)
\simeq
\frac{T}{8\pi^{3/2}}
\int_0^\infty dv\,\mathcal{P}(v)\,
\frac{1}{v}\,
\frac{1}{k}\,
{\rm erf}\!\left(
\frac{k/k_v}{\sqrt{1+(k/k_\sigma)^2}}
\right),
\label{eq:J_k_def_recoil_free}
\end{align}
up to corrections of order $\mathcal O(k^2/k_*^2)$.
Here $k_v:=1/(vT)$ is the Doppler scale associated with the finite observation time.

For fixed velocity $v$, the kernel approaches a constant in the deep infrared, where the linear behavior of the error function cancels the explicit factor $1/k$.
As $k$ increases, a $1/k$ tail develops once the argument of the error function becomes of order unity.
When $k_v\ll k_\sigma$, the transition occurs at $k\sim k_v$, and the kernel immediately acquires $1/k$ behavior.
When $k_\sigma\ll k_v$, a formal $1/k$ behavior appears already for $k\gtrsim k_\sigma$, but its coefficient is suppressed by the small ratio $k_\sigma/k_v$.
Consequently, the dominant contribution arises from momenta $k\gtrsim k_v$.

For the thermal gas model specified in Sec.~\ref{subsec:decoherence_spherical}, one finds
\begin{align}
k_\sigma
=
\sqrt{\frac{4}{3\pi}}\,
k_{v,{\rm typ}},
\qquad
k_{v,{\rm typ}}
:=
\frac1{v_{\rm typ}T},
\label{eq:ksigma_kv_relation}
\end{align}
so that the two scales differ only by an order-unity factor.
Hence, distinguishing them does not affect the leading order estimates. 
For notational simplicity, we use $k_\sigma$ to denote the infrared scale in the following.
Then the kernel exhibits the asymptotic behavior
\begin{align}
\mathcal J(k)
\simeq
\begin{cases}
J_0,
& k\ll k_\sigma,
\\[6pt]
\dfrac{T}{8\pi^{3/2}}
\dfrac{A_\infty}{k},
& k_\sigma\ll k\ll k_*,
\end{cases}
\label{eq:J_piecewise}
\end{align}
where
$
J_0:=T^2/(4\pi^2)
$
and
\begin{align}
A_\infty
:=
\int_0^\infty dv\,\mathcal P(v)\,
\frac{1}{v}\,
{\rm erf}\!\left(
\frac{mv}{\sqrt{2}\sigma_k}
\right)
= \frac{2}{\pi v_{\rm typ}}.
\label{eq:A_infty_def}
\end{align}
The final expression follows from Eq.~\eqref{eq:MB_distribution} together with the Gaussian--error-function identity Eq.~\eqref{eq:Gaussian_erf_identity}.

For finite radius $r$, the lower limit of the diagonal integral is set by $r^{-1}$.
A non-negligible contribution is therefore present only when $r^{-1}<k_{\rm UV}$.
Combining the diagonal condition with the asymptotic form Eq.~\eqref{eq:J_piecewise}, the infrared cutoff scale of the integral is
\begin{align}
k_{\rm IR}(r)
:=
\max(k_\sigma,r^{-1}).
\label{eq:kIR_def}
\end{align}
Hence, when $k_{\rm IR}(r)<k_{\rm UV}$, the diagonal contribution becomes
\begin{align}
\Gamma^{\rm diag}_{<r}
\simeq
\mathcal C
\frac{T A_\infty}{48\pi^{3/2}}
\Delta x^2
\ln\!\left(
\frac{k_{\rm UV}}{k_{\rm IR}(r)}
\right),
\label{eq:Gamma_diag_log_r}
\end{align}
which depends on the cutoff momenta only logarithmically.
Equivalently,
\begin{align}
\Gamma^{\rm diag}_{<r}
\simeq
\Gamma_0
\begin{cases}
0,
& r\ll k_{\rm UV}^{-1},
\\[4pt]
\ln(k_{\rm UV}r),
& k_{\rm UV}^{-1}\ll r\ll k_\sigma^{-1},
\\[4pt]
\ln(k_{\rm UV}/k_\sigma),
& r\gg k_\sigma^{-1},
\end{cases}
\label{eq:Gamma_diag_log_piecewise}
\end{align}
where
\begin{align}
\Gamma_0
:=
\mathcal C
\frac{T A_\infty}{48\pi^{3/2}}
\Delta x^2 .
\label{eq:Gamma0_def}
\end{align}
Thus the diagonal contribution accumulates over the radial interval $k_{\rm UV}^{-1}\lesssim r\lesssim k_\sigma^{-1}$ logarithmically and saturates for $r\gtrsim k_\sigma^{-1}$.
Formally, taking the observation time to infinity drives $k_\sigma$ to zero and would lead to a formal logarithmic divergence within the effective dipole description.
In a realistic setup, however, the dipole description is expected to break down at sufficiently large distances, where the combined system--apparatus configuration crosses over to a quadrupole structure.
This introduces an additional infrared cutoff and removes the formal divergence.

The result may also contain finite non-logarithmic pieces.
When $r^{-1}<k_\sigma$, the infrared part of the diagonal integral for $r^{-1}\lesssim k\lesssim k_\sigma$ gives a contribution $\mathcal C J_0\Delta x^2(k_\sigma-r^{-1})/6$.
If $k_\sigma\ll k_\Delta\ll k_{R_s}$, there is also a finite contribution from $k_\Delta\lesssim k\lesssim k_{R_s}$, given by $3\Gamma_0$.
These terms are not enhanced by $\ln(k_{\rm UV}/k_\sigma)$ and are subleading under the hierarchy $k_\sigma\ll k_{\rm UV}$.

\subsection{Off-diagonal sector}
\label{subsec:decoherence_nondiag}
We now evaluate the off-diagonal contribution defined by Eq.~\eqref{eq:Gamma_off_def}.
The diagonal contribution is absent for $r\ll k_{\rm UV}^{-1}$.
In this regime, the kernel momenta satisfy $k,\ k' \lesssim k_{\rm UV}\ll r^{-1}$.
Since $\tilde\varphi_r(\bm q)$ varies on the momentum scale $r^{-1}$, the window factors in Eq.~\eqref{eq:Ir_window_def} can be expanded in the kernel momenta,
\begin{align}
\tilde\varphi_r(\bm q-\bm k)
=
\tilde\varphi_r(\bm q)
-
k_i\frac{\partial}{\partial q_i} \tilde\varphi_r(\bm q)
+
\frac12 k_i k_j\frac{\partial^2}{\partial q_i\partial q_j} \tilde\varphi_r(\bm q)
+\cdots.
\end{align}
Substituting this expansion into Eq.~\eqref{eq:Ir_window_def}, rotational symmetry gives
\begin{align}
I_r(\bm k,\bm k')
=
I_0
+
(2\pi)^3 A_r(k^2+k'^2)
+
(2\pi)^3 B_r\,\bm k\cdot\bm k'
+\cdots,
\label{eq:Ir_small_r_expansion_full}
\end{align}
where the coefficients are given by
\begin{align}
I_0
&:=
(2\pi)^3
\int d^3q\;
\tilde\varphi_r^2(\bm q)\,
\mathcal J(q),
\\
A_r
&:=
\frac{1}{6}
\int d^3q\;
\tilde\varphi_r(\bm q)\,
\nabla_q^2\tilde\varphi_r(\bm q)\,
\mathcal J(q),
\\
B_r
&:=
\frac{1}{3}
\int d^3q\;
\left(\nabla_q\tilde\varphi_r(\bm q)\right)^2\,
\mathcal J(q).
\label{eq:I0_Ar_Br_defs}
\end{align}

Substituting the expansion Eq.~\eqref{eq:Ir_small_r_expansion_full} into Eq.~\eqref{eq:Gamma_window_spatial}, we also expand the dipole factors for $k,k'\lesssim k_{\rm UV}$,
\begin{align}
\sin\!\left(
\frac{\bm k\cdot\Delta\bm x}{2}
\right)
\simeq
\frac{\bm k\cdot\Delta\bm x}{2},
\end{align}
and similarly for $\bm k'$.
The terms proportional to $I_0$ and $A_r$ then vanish after angular integration due to rotational symmetry, while the cross term proportional to $\bm k\cdot\bm k'$ gives the leading off-diagonal contribution
\begin{align}
\Gamma^{\rm off}_{<r}
&\simeq
\frac{(2\pi)^3\kappa^4}{32}
M^2m^2 n\,
B_r\,
\mathcal S_{\rm dip}(\Delta x,R_s).
\label{eq:Gamma_off_small_general}
\end{align}
Here 
\begin{align}
\mathcal S_{\rm dip}(\Delta x,R_s)
&:=
\left|
\int\frac{d^3k}{(2\pi)^3}
e^{-R_s^2k^2/2}
\frac{\bm k}{k^2}
\frac{\bm k\cdot\Delta\bm x}{2}
\right|^2
\nonumber\\
&=
\frac{\Delta x^2}{288\pi^3R_s^6},
\label{eq:S_dip_eval}
\end{align}
where the second line follows from rotational symmetry.

We evaluate $B_r$ in the small-window regime $r\ll k_{\rm UV}^{-1}$.
Since $(\partial/\partial q_i)\tilde\varphi_r(\bm q)\propto F'(qr)$, the integral in Eq.~\eqref{eq:I0_Ar_Br_defs} is dominated by momenta $q\sim r^{-1}$.
Using $r^{-1}\gg k_\sigma$, the dominant region therefore satisfies $q\gg k_\sigma$, and we may use the intermediate asymptotic form for $k_\sigma \ll q \ll k_*$ in Eq.~\eqref{eq:J_piecewise}.
Substituting this into Eq.~\eqref{eq:I0_Ar_Br_defs}, and using $\int_0^\infty dx\,x[F'(x)]^2=3/4$, we obtain
\begin{align}
B_r
=
\frac{2}{9}\pi^{3/2}
T A_\infty r^6 .
\label{eq:Br_eval}
\end{align}

Substituting Eq.~\eqref{eq:S_dip_eval} and Eq.~\eqref{eq:Br_eval} into Eq.~\eqref{eq:Gamma_off_small_general}, we obtain
\begin{align}
\Gamma^{\rm off}_{<r}
&\simeq
\frac{\sqrt\pi}{81}
\frac{\kappa^4}{32}
M^2m^2 n\,
\frac{T}{v_{\rm typ}}
\Delta x^2
\frac{r^6}{R_s^6}.
\label{eq:small_r_result_MB}
\end{align}
Equation~\eqref{eq:small_r_result_MB} shows that the off-diagonal contribution in the small-window regime grows only as a power law, $\Gamma^{\rm off}_{<r}\propto r^6$.
Thus, the cumulative decoherence function can be evaluated analytically for $r\lesssim k_{\rm UV}^{-1}$ before the logarithmic diagonal contribution sets in.
For $r\gtrsim k_{\rm UV}^{-1}$, the expansion in external momenta is no longer valid, and the off-diagonal contribution must be evaluated from Eq.~\eqref{eq:Ir_window_def} without approximation.

\subsection{Estimate of the decoherence function}
\label{subsec:decoherence_rate}

We now estimate the decoherence function for the thermal gas by evaluating the large-radius limit of the cumulative decoherence function.
We focus on the recoil-free case with $k_\sigma\ll k_{\rm UV}\ll k_*$ and under the hierarchy $k_\sigma\ll k_{\rm UV}$. 
In this regime, the dominant contribution is given by the large-radius limit of Eq.~\eqref{eq:Gamma_diag_log_piecewise}.
Substituting Eq.~\eqref{eq:Gamma0_def} with Eqs.~\eqref{eq:C_prefactor} and \eqref{eq:A_infty_def}, and restoring the suppressed factor of $\hbar$ for the SI estimate, we obtain
\begin{align}
\Gamma
&\simeq
\frac{\pi^{3/2}\kappa^4}{48\hbar^2}
M^2m^2 n
\frac{T}{v_{\rm typ}}
\Delta x^2
\ln\!\left(
\frac{k_{\rm UV}}{k_\sigma}
\right).
\label{eq:Gamma_lead_MB}
\end{align}

As a conservative reference environment, we consider molecular nitrogen at room temperature and atmospheric pressure, with $m = 28\,{\rm amu}\simeq 4.65\times10^{-26}\ {\rm kg},\ T_{\rm gas} = 300\ {\rm K}$, and $P_{\rm gas} = 1\ {\rm atm}\simeq 1.0\times10^5\ {\rm Pa}$.
As a representative benchmark motivated by levitated particles, we take $M = 10^{-17}\ {\rm kg},\ R_s = 10^{-7}\ {\rm m},\ \Delta x = 10^{-8}\ {\rm m}$.
For these parameters we have $k_{R_s}\ll k_*$ and $k_{R_s}\ll k_\Delta$, so that the recoil-free condition is well satisfied and $k_{\rm UV}\simeq k_{R_s}$.
With these choices, Eq.~\eqref{eq:Gamma_lead_MB} gives
\begin{align}
\Gamma
&=
1.2\times 10^{-26}\,
\left(\frac{M}{10^{-17}\ {\rm kg}}\right)^2
\left(\frac{m}{4.65\times10^{-26}\ {\rm kg}}\right)^2
\left(\frac{n}{2.4\times10^{25}\ {\rm m}^{-3}}\right)
\left(\frac{v_{\rm typ}}{4.8\times10^2\ {\rm m/s}}\right)^{-1}
\nonumber\\
&\qquad\times
\left(\frac{T}{1\ {\rm s}}\right)
\left(\frac{\Delta x}{10^{-8}\ {\rm m}}\right)^2
\frac{
\ln\!\left(k_{\rm UV}/k_\sigma\right)
}{
\ln\!\left(k_{{\rm UV},{\rm ref}}/k_{\sigma,{\rm ref}}\right)
}\,.
\label{eq:Gamma_benchmark_nvt}
\end{align}
Here $k_{{\rm UV},{\rm ref}}$ and $k_{\sigma,{\rm ref}}$ denote the corresponding values of $k_{\rm UV}$ and $k_\sigma$ evaluated for the reference parameters introduced above.

\medskip
To visualize the parametric dependence of the decoherence function, we numerically evaluate $\Gamma$ over representative regions of the parameter space.
The overlaid experimental regions are schematic envelopes motivated by representative values for matter-wave interferometers, levitated particles, and optomechanical systems.
For matter-wave interferometers, the mass and size ranges extend from individual atomic species to large-molecule interferometry~\cite{Cronin:2009zz,Gerlich:2011kpf}, while the superposition scale ranges from typical laboratory interferometers to large-baseline experiments with demonstrated wave-packet separations approaching the meter scale~\cite{Cronin:2009zz,Kovachy:2015xcp}.
For levitated particles, the mass and size ranges extend from nanometer-scale dielectric particles to larger levitated dielectric systems proposed for tests of gravity-mediated entanglement~\cite{Bateman:2013zna,Bose:2017nin}, with superposition scales ranging from nanometer-scale motional delocalization to spatial separations reaching the micrometer regime~\cite{Romero-Isart:2011sdw,Bose:2017nin}.
For optomechanical systems, the mass and size ranges extend from femtogram-scale nanobeam and photonic-crystal resonators to mesoscopic mechanical devices~\cite{Eichenfield:2009jij,Marshall:2002exi}, while the displacement scale ranges from the zero-point motion of high-frequency resonators to picometer-scale mechanical displacements~\cite{Teufel:2011smx,Aspelmeyer:2013lha}.

\begin{figure}
\centering
\includegraphics[width=0.85\linewidth]{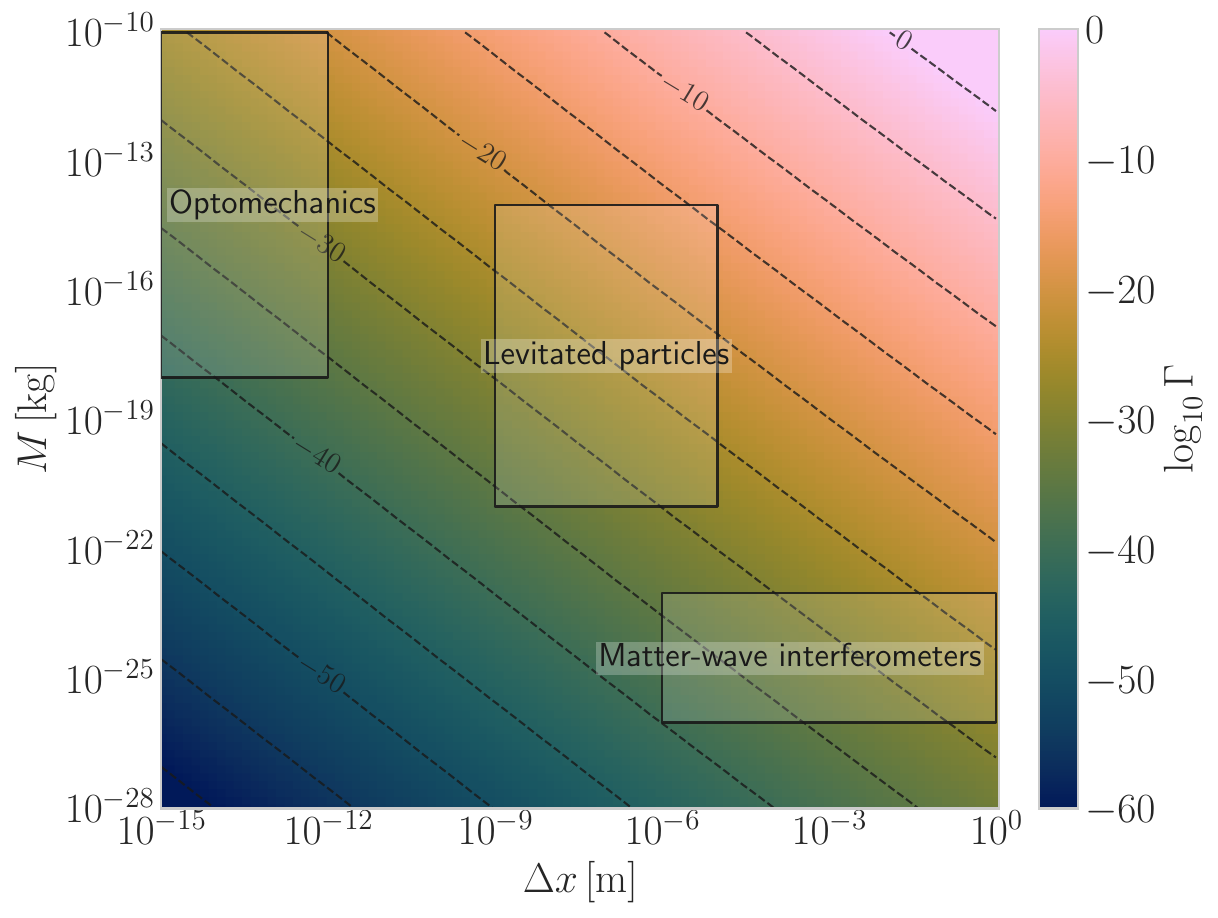}
\caption{
Contour plot of $\log_{10}\Gamma$ in the $(\Delta x, M)$ plane, obtained by numerically evaluating the saturated limit $r\to\infty$ of Eq.~\eqref{eq:Gamma_1D_diag_r} for a dilute thermal gas with $m = 28\,{\rm amu}\simeq 4.65\times10^{-26}\ {\rm kg},\ T_{\rm gas} = 300\ {\rm K},\ P_{\rm gas} = 1\ {\rm atm}\simeq 1.0\times10^{5}\ {\rm Pa}$.
Dashed lines indicate constant values of $\log_{10}\Gamma$, and shaded regions show its magnitude.
Representative experimental regimes are overlaid: matter-wave interferometers, levitated particles, and optomechanical systems.
The system size is fixed to $R_s=10^{-7}\,{\rm m}$, and the resulting dependence on $R_s$ is weak.
}
\label{fig:Gamma_contour}
\end{figure}

Figure~\ref{fig:Gamma_contour} shows $\log_{10}\Gamma$ on the $(\Delta x, M)$ plane, obtained by numerically evaluating Eq.~\eqref{eq:Gamma_1D_diag_r} in the limit $r\to\infty$ for the reference gas parameters introduced above and a fixed system size $R_s=10^{-7}\,{\rm m}$.
The contours are approximately linear in the log--log plane, reflecting the dominant scaling $\Gamma \propto M^2 \Delta x^2$ with a weak logarithmic correction.
The dependence on the internal size $R_s$ is weak, because $R_s$ enters only through the ultraviolet cutoff $k_{\rm UV}$ inside the logarithm.
As an example of an aggressive parameter regime designed to probe the quantum nature of gravity, one may consider proposals that aim to detect entanglement generated by the mutual gravitational interaction between massive objects prepared in spatial superpositions~\cite{Bose:2017nin}.
The representative parameters are roughly $M\simeq10^{-14}\,{\rm kg}$, $R_s\simeq1\,\mu{\rm m}$, and $\Delta x\simeq250\,\mu{\rm m}$, corresponding to the upper-right edge of the levitated-particle region in Fig.~\ref{fig:Gamma_contour}.
Even for such large masses and macroscopic superposition sizes, the gravitationally induced decoherence remains very small $\sim 10^{-13}$, despite the relatively large environmental density corresponding to atmospheric pressure.

\begin{figure*}[t]
\centering

\begin{minipage}[t]{0.48\textwidth}
    \centering
    \includegraphics[width=\linewidth]{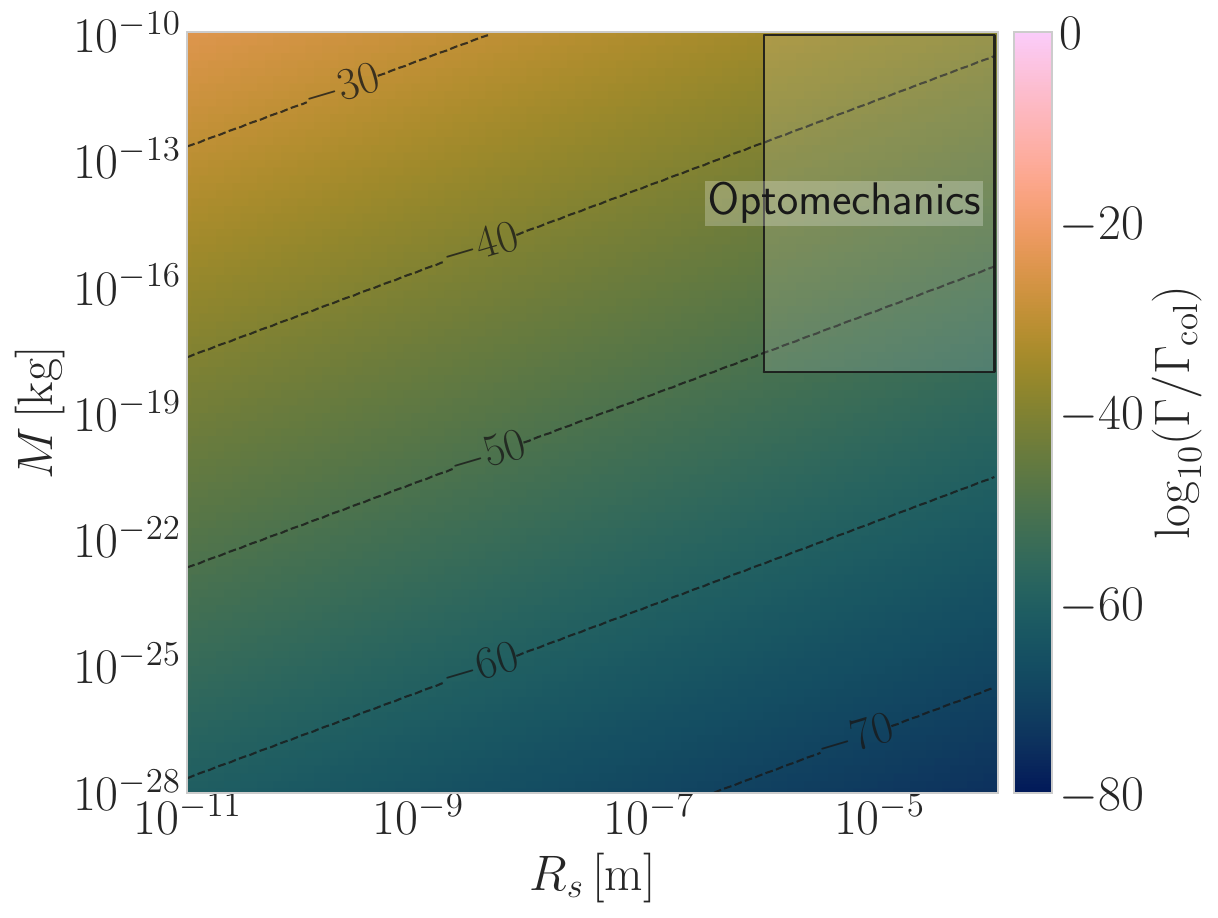}\\[0.3em]
    {\small (a) $\Delta x = 10^{-13}\,\mathrm{m}$}
\end{minipage}
\hfill
\begin{minipage}[t]{0.48\textwidth}
    \centering
    \includegraphics[width=\linewidth]{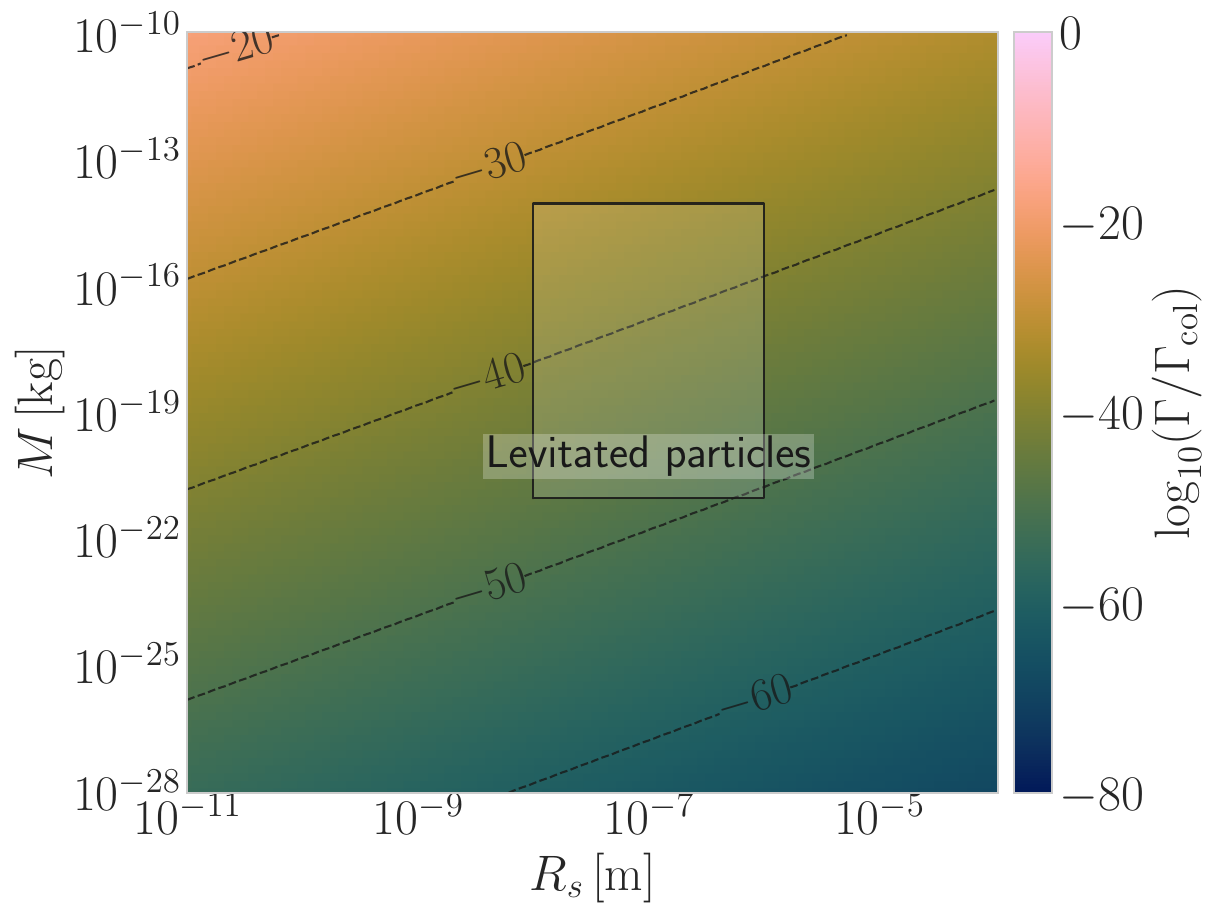}\\[0.3em]
    {\small (b) $\Delta x = 10^{-8}\,\mathrm{m}$}
\end{minipage}

\vspace{0.8em}

\begin{minipage}[t]{0.48\textwidth}
    \centering
    \includegraphics[width=\linewidth]{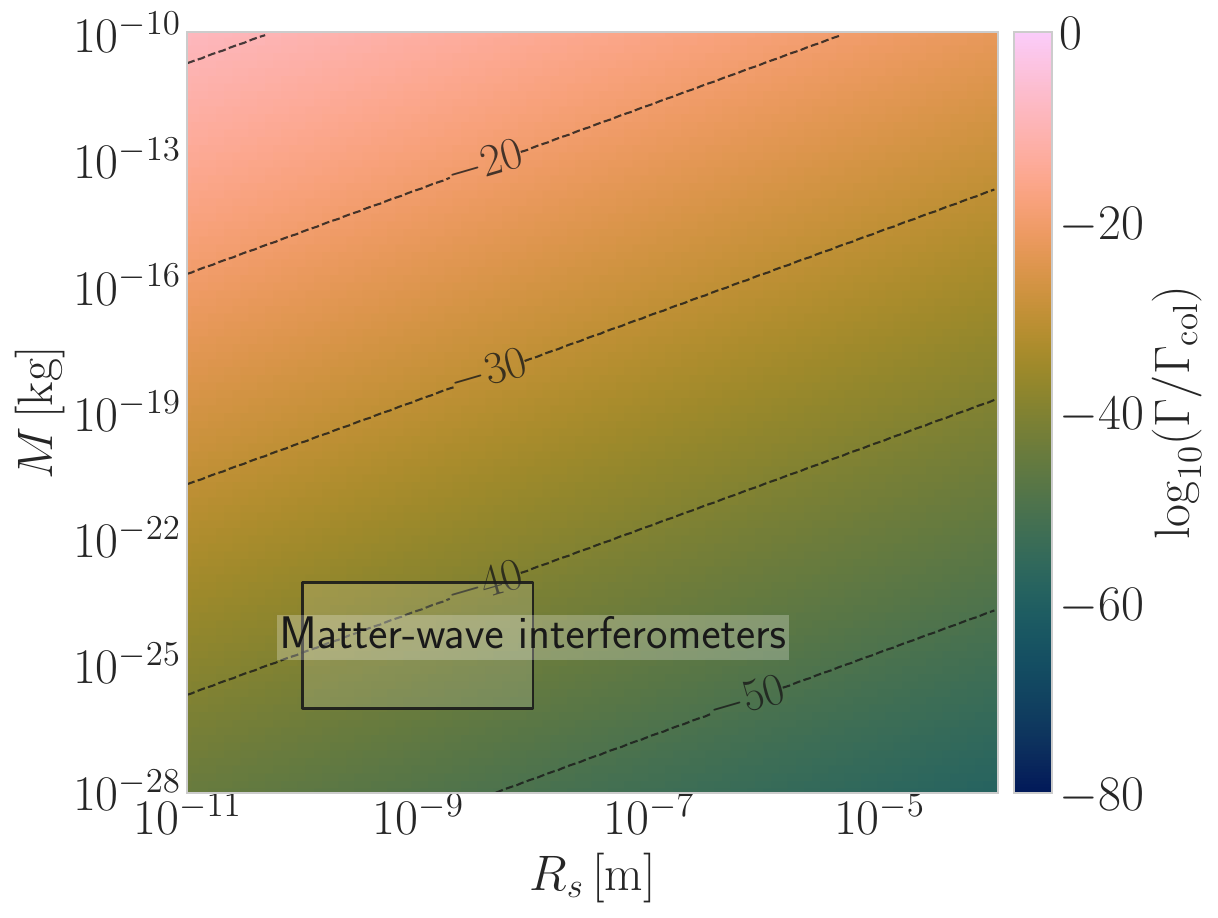}\\[0.3em]
    {\small (c) $\Delta x = 10^{-3}\,\mathrm{m}$}
\end{minipage}

\caption{
Contour plots of $\log_{10}(\Gamma/\Gamma_{\rm col})$ in the $(R_s,M)$ plane for fixed superposition sizes:
(a) $\Delta x = 10^{-13}\,\mathrm{m}$,
(b) $\Delta x = 10^{-8}\,\mathrm{m}$, and
(c) $\Delta x = 10^{-3}\,\mathrm{m}$.
Here $\Gamma$ is obtained by numerically evaluating the saturated limit $r\to\infty$ of Eq.~\eqref{eq:Gamma_1D_diag_r}, while $\Gamma_{\rm col}$ is estimated from Eq.~\eqref{eq:Gamma_col} for a dilute thermal gas with
$m = 28\,{\rm amu}\simeq 4.65\times10^{-26}\,{\rm kg}$,
$T_{\rm gas}=300\,{\rm K}$, and
$P_{\rm gas}=1\,{\rm atm}\simeq1.0\times10^{5}\,{\rm Pa}$.
Dashed lines indicate constant values of $\log_{10}(\Gamma/\Gamma_{\rm col})$.
In each panel, only the representative experimental regime relevant to the chosen value of $\Delta x$ is overlaid schematically.
}
\label{fig:ratio_three_panels}
\end{figure*}

Figure~\ref{fig:ratio_three_panels} shows the ratio $\log_{10}(\Gamma/\Gamma_{\rm col})$ in the $(R_s,M)$ plane for three fixed values of the superposition size, $\Delta x=10^{-13}\,\mathrm{m}$, $10^{-8}\,\mathrm{m}$, and $10^{-3}\,\mathrm{m}$ for the reference gas parameters introduced above.
Here, $\Gamma$ is obtained by numerically evaluating the same one-dimensional integral~\eqref{eq:Gamma_1D_diag_r}, while the collisional decoherence function $\Gamma_{\rm col}$ is estimated using the infinite-mass limit with an isotropic geometric cross section $\sigma=\pi R_s^2$~\cite{Hornberger:2003umw},
\begin{align}
\Gamma_{\rm col}
=
T\,n\,\pi R_s^2\,v_{\rm typ}
\left(
1-\frac{D(\xi)}{\xi}
\right),
\qquad
\xi
=
\frac{\sqrt{2mk_B T_{\rm gas}}}{\hbar}\,\Delta x,
\label{eq:Gamma_col}
\end{align}
where $D(\xi):=e^{-\xi^2}\int_0^\xi e^{t^2}dt$ is the Dawson function.
Across the representative parameter ranges, the ratio remains negative, showing that the collisional decoherence dominates over the gravitational contribution.
Although the relative importance of gravitational decoherence increases with $\Delta x$, reflecting the scaling $\Gamma\propto M^2\Delta x^2$, it remains subdominant in the experimentally relevant regions considered here.
This demonstrates that gravitationally induced decoherence in a dilute thermal gas is generically negligible compared with standard collisional decoherence in current experimental settings.

\section{Conclusion}
\label{sec:conclusion}

In this paper we formulated the gravitationally induced decoherence of a massive object prepared in a spatial superposition by combining linearized gravity with the closed-time-path influence-functional formalism. 
After integrating out the metric perturbation and the environmental scalar field, we obtained a general decoherence functional written as a bilinear form of the system stress--energy tensor difference and an effective gravitationally dressed noise kernel. 
We then specialized this framework to a dilute nonrelativistic thermal gas, introduced finite-time and finite-size coarse graining, and formulated a cumulative spatial-window decoherence function. 
Its diagonal contribution reduces to a tractable one-dimensional momentum integral, whose large-radius limit gives a convergent decoherence function $\Gamma$.

In the recoil-free regime relevant to realistic laboratory parameters, the dominant contribution arises from an intermediate momentum region. 
This yields a leading behavior, $\Gamma \propto M^2\Delta x^2\ln(k_{\rm UV}/k_\sigma)$, showing that the decoherence grows with the square of the system mass $M$ and branch separation $\Delta x$, while depending only logarithmically on the ratio between the ultraviolet cutoff scale $k_{\rm UV}$ set by the source size $R_s$ and superposition scale $\Delta x$, and the environmental infrared cutoff scale $k_\sigma$.
The numerical evaluation on the $(\Delta x,M)$ plane confirms this structure and the subdominance of the gravitational decoherence induced by a residual gas, throughout experimentally relevant regions, including matter-wave interferometers, levitated particles, and optomechanical systems.
Furthermore, by varying the spatial support of the environment, we have shown that this logarithmic contribution is accumulated over a wide range of radii, $k_{\rm UV}^{-1}\lesssim r \lesssim k_\sigma^{-1}$, while regions at smaller distances give only parametrically suppressed contributions.

The present analysis was carried out using an effective dipole description of the branch separation.
For a physical mass distribution satisfying momentum conservation, the leading far-field contribution is expected to begin at quadrupole order once the surrounding apparatus is included to cancel the net dipole moment.
The dipole approximation adopted here therefore provides a conservative benchmark for the gravitational decoherence effect.
In a realistic setup, the finite size of the combined system--apparatus configuration is expected to introduce an additional IR cutoff, further reducing the decoherence rate.

To conclude, despite its long-range and unscreenable nature, the gravitational contribution from a dilute gas is generically negligible compared to standard collisional decoherence under realistic conditions.
At the same time, the present framework provides a first-principles description of gravitational quantum decoherence induced by environmental matter fluctuations and can be extended to more general environments.

\section*{Acknowledgements}
H.T. is supported by the Hakubi project at Kyoto University, and by Japan Society for the Promotion of Science (JSPS) KAKENHI Grant No. JP22K14037 and No. JP26K17146.
S.T. is supported by the Research Fellow program of Kyoto University.
T.T. is supported by JSPS KAKENHI Grant No. JP23H00110. 
This work is also supported by SPIRIT2 2026 of Kyoto University.


\bibliography{bib}

\end{document}